\documentclass[12pt]{article}
\usepackage{amsmath}
\usepackage{graphicx} %,psfrag,epsf}
\usepackage{enumerate}
\usepackage{natbib}
\usepackage{url} % not crucial - just used below for the URL 
\usepackage{float}
\usepackage{pdflscape}
\usepackage{adjustbox}
\usepackage{booktabs} 
\usepackage{hyperref}

\newcommand{\blind}{0}

\begin{document}

\def\spacingset#1{\renewcommand{\baselinestretch}%
{#1}\small\normalsize} \spacingset{1}

%%%%%%%%%%%%%%%%%%%%%%%%%%%%%%%%%%%%%%%%%%%%%%%%%%%%%%%%%%%%%%%%%%%%%%%%%%%%%%

\if0\blind
{
  \title{\bf Comparison of methods for mediation analysis with multiple correlated mediators}
  \author{Mary Appah\thanks{
    This research project is supported by cooperative agreement U01 NS041588 co-funded by the National Institute of Neurological Disorders and Stroke (NINDS) and the National Institute on Aging (NIA), National Institutes of Health, Department of Health and Human Service. The content is solely the responsibility of the authors and does not necessarily represent the official views of the NINDS or the NIA.  The authors thank the other investigators, the staff, and the participants of the REGARDS study for their valuable contributions. A full list of participating REGARDS investigators and institutions can be found at: \href{https://www.uab.edu/soph/regardsstudy/}{https://www.uab.edu/soph/regardsstudy/}}\hspace{.2cm}\\
    Department of Biostatistics, University of Alabama at Birmingham,\\
         D. Leann Long  \\
    Department of Biostatistics and Data Science, Wake Forest University\\
   George Howard\\
   Department of Biostatistics, University of Alabama at Birmingham\\
   Melissa J. Smith\\
   Department of Biostatistics, University of Alabama at Birmingham}
  \maketitle
} \fi

\if1\blind
{
  \bigskip
  \bigskip
  \bigskip
  \begin{center}
    {\LARGE\bf Title}
\end{center}
  \medskip
} \fi

\bigskip
\begin{abstract}
Various methods have emerged for conducting mediation analyses with multiple correlated mediators, each with distinct strengths and limitations. However, a comparative evaluation of these methods is lacking, providing the motivation for this paper. This study examines six mediation analysis methods for multiple correlated mediators that provide insights to the contributors for health disparities. We assessed the performance of each method in identifying joint or path-specific mediation effects in the context of binary outcome variables varying mediator types and levels of residual correlation between mediators. Through comprehensive simulations, the performance of six methods in estimating joint and/or path-specific mediation effects was assessed rigorously using a variety of metrics including bias, mean squared error, coverage and width of the 95$\%$ confidence intervals.  Subsequently, these methods were applied to the REasons for Geographic And Racial Differences in Stroke (REGARDS) study, where differing conclusions were obtained depending on the mediation method employed. This evaluation provides valuable guidance for researchers grappling with complex multi-mediator scenarios, enabling them to select an optimal mediation method for their research question and dataset.
\end{abstract}

\noindent%
{\it Keywords:} Method Comparison, Mediation Analysis, Multiple Correlated Mediators
\vfill

\newpage
\spacingset{1.45} % DON'T change the spacing!
\section{Introduction}
\label{sec:intro}

In health research, understanding the pathways through which exposures influence outcomes provides a basis for targeting interventions. Mediation analysis, a guiding light in this field, decomposes the total effect of an exposure on an outcome into direct and indirect effects, where the factors in the indirect path reflect potential causal pathways. While traditionally focused on dissecting the role of a single mediator, there’s growing interest in unraveling both joint indirect effects through a set of mediators and path-specific indirect effects through specific mediators. For instance, researchers may be interested in deciphering how various lifestyle factors including diet, smoking, and physical activity collectively and individually explain the observed disparities in health outcomes. Conducting mediation analyses with multiple mediators poses unique challenges, including high correlation and interactions between potential mediators, necessitating specialized statistical methods. Moreover, variables vary in form, from continuous measures like body mass index to binary indicators like smoking status. This diversity necessitates statistical tools that can seamlessly navigate through these difficulties.

In a scoping review by \cite{rijnhart2021mediation}, it was found that more than 30\% of included papers from 2015 to 2019 employed mediation analysis with multiple mediators. Mediation models fall into two broad classes: traditional mediation, which often follows the framework proposed by \cite{BaronKenny1986} and focuses on partitioning the total effect into direct and indirect effects, and causal mediation, which uses a counterfactual-based approach to identify and estimate causal pathways and account for confounding variables. Most of these papers used a traditional mediation framework rather than a causal mediation framework. Notably, the REasons for Geographic and Racial Differences in Stroke (REGARDS) Study \citep{howard2005reasons} and related research have utilized various methodologies for multiple mediator analysis to assess contributors to health disparities, serving as the motivation for this work. For instance, \cite{tajeu2020black} utilized the Inverse Odds Ratio Weighting (IORW) method \citep{TchetgenTchetgen2013} to assess the joint indirect effect of numerous mediators to assess the joint indirect effect of numerous mediators to evaluate contributors to racial disparities in cardiovascular disease mortality, whereas \cite{carson2021sex} utilized the difference-in-coefficients method \citep{BaronKenny1986} to assess the joint indirect effect of individual and neighborhood factors that contribute to the racial disparity in diabetes incidence. Similarly, \cite{howard2018association} employed the difference-in-coefficients method \citep{BaronKenny1986} 1986) to study the contributors to racial disparities in incident hypertension.

In recent years, several statistical methods have emerged for mediation analyses with multiple correlated and contemporaneous mediator variables  \citep{TchetgenTchetgen2013, Wang2013EstimationOC, vanderweele2014mediation, lange2014assessing, nguyen2016causal, yu2017mma, taguri2018causal, kim2019bayesian, jerolon2021causal, roy2024bayesian}. Each method has its strengths and limitations, emphasizing the importance of selecting the most suitable method for a given research question and dataset. For example, while methods by \cite{TchetgenTchetgen2013} and \cite{vanderweele2014mediation}  estimate joint indirect effects, others like \citep{Wang2013EstimationOC, lange2014assessing, yu2017mma, taguri2018causal, kim2019bayesian, jerolon2021causal, roy2024bayesian}  were developed to identify path-specific indirect effects, albeit requiring additional causal assumptions. Some methods, such as those by \citep{Wang2013EstimationOC, kim2019bayesian, jerolon2021causal}, account for residual correlation between mediator variables, while others, like \citep{lange2014assessing, taguri2018causal}, assume mediator independence. Moreover, Bayesian non-parametric methods \citep[e.g.][]{kim2019bayesian, roy2024bayesian} offer flexibility but are more computationally burdensome compared to frequentist regression-based approaches  \citep[e.g.][]{TchetgenTchetgen2013, vanderweele2014mediation, lange2014assessing}.

Some studies have compared one or two of these methods under specific conditions, revealing differences in performance based on mediator type, correlation levels, and other factors. For instance, \cite{jerolon2021causal} compared their method to a simple approach of analyzing the mediators individually, as well as to the regression-based approach proposed by \cite{vanderweele2014mediation}. The simple mediation approach involves assessing each mediator separately, focusing on one at a time. However, it has been shown that analyzing mediators independently can lead to biased results, especially if the mediators are correlated \citep{vanderweele2014mediation}. Additionally, they compared the path-specific effects of their method to those of  \cite{vanderweele2014mediation}. This comparison may be likened to contrasting an appropriate method with an inappropriate one, as the method proposed by \cite{vanderweele2014mediation} does not estimate path-specific effects and the authors explained that analyzing mediators independently when the mediators are correlated would lead to biased results.

Despite the surge in interest surrounding mediation analyses with multiple correlated mediators, there exists a noticeable void in the literature; a lack of systematic comparisons between the various methodologies. To our knowledge, no study has undertaken a meticulous comparison and contrast of each of these mediation analysis approaches, particularly in the context of multiple correlated and potentially interacting mediators. Our focus extends beyond mere comparison; we aim to assess these methods in studies involving binary health outcomes and a mix of mediator variable types. To achieve this, we perform both an in-depth simulation study and an analysis of the REGARDS dataset. Through these efforts, we aspire to offer a roadmap for statistical analysts, unraveling the intricacies of each method and providing a foundation for informed decision-making in the realm of mediation analysis.

In this study, we compare the performance of six methods out of this diverse pool of existing methods for multiple correlated and contemporaneous mediator variables. Selection criteria for inclusion of an approach for mediation assessment for this comparative assessment were: 1) a low-to-moderate computation time in terms of computing the causal effect of interest, 2) accounting for residual correlation between the mediators if computing path-specific effects, 3) adjustment for confounders, and 4) readily available software. The methods selected for comparison in this paper are those developed by  \cite{BaronKenny1986}, \cite{TchetgenTchetgen2013}, \cite{Wang2013EstimationOC}, \cite{vanderweele2014mediation}, and \cite{jerolon2021causal}.

The remainder of this paper is organized as follows. In Section 2, we introduce the general and counterfactual notation as well as the key assumptions for each of the six approaches for mediation analysis, and describe the methods of each approach being compared in our simulation study. In section 3 we present results from the simulation studies to comparing the methods described in the previous sections. In Section 4 we illustrate the application of these methods by determining whether a set of correlated health behavior variables will mediate the relationship between baseline household income and hypertension incidence in the REGARDS study. We conclude with a discussion of the strengths and limitations of these methods and recommendations for statistical practice in Section 5.

\section{Methods for mediation analysis with multiple mediators}
\label{sec:meth}
\begin{figure}[h]
\centering
\includegraphics[scale = 0.7]{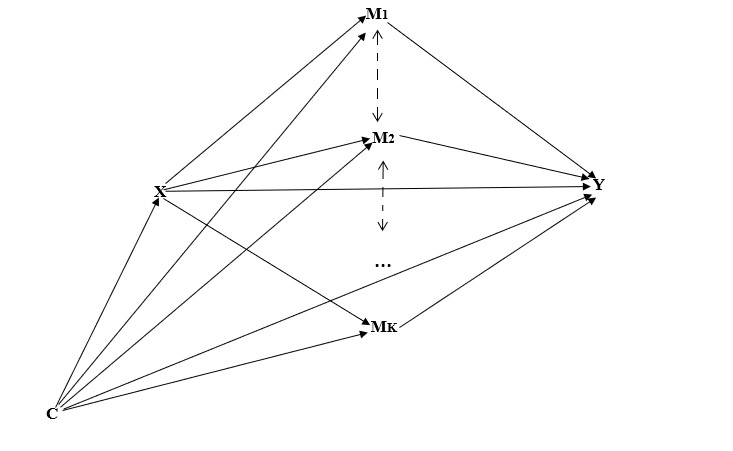}
\caption{A Directed Acyclic Graph (DAG) showing the relationship between variables in mediation analysis with multiple correlated mediators}
\label{Figure:dag}
\end{figure}

We begin by introducing our notation for describing the methods used in this study. Let $Y$ represent a binary outcome variable $(Y \in \{0,1\})$, $X$ be a binary exposure variable $(X \in \{0,1\})$, $\boldsymbol{C}$ be a vector of baseline covariates that may affect the mediator, exposure, and/or outcome, and $\boldsymbol{M}= [M_1, M_2, ..., M_K]^{'}$ be a vector of an individual's $K$ observed correlated mediator values that may be on the pathway from exposure to outcome. The causal diagram in Figure \ref{Figure:dag} illustrates the presumed causal relationships between these variables defined above. We focus on the setting where mediators are not causally related \citep[see][]{Wang2013EstimationOC, kim2019bayesian, jerolon2021causal} and therefore contemporaneous in nature rather than ordered \citep[see][]{daniel2015causal, gao2019bayesian}. This could be due to an underlying common cause that induces a correlation between mediators, as indicated by the dotted bidirectional arrows in Figure \ref{Figure:dag}. Without loss of generality, we will consider the setting where we have two correlated mediator variables in the remainder of this paper. That is, $\boldsymbol{M}= [M_1, M_2]^{'}$  will denote the vector of mediators for each individual, with $M_1$ representing the first mediator and $M_2$ representing the second mediator.

Many researchers have extended the traditional methods originally developed by \cite{BaronKenny1986} to a causal inference framework that allows for numerous link functions, exposure-mediator interactions, and even mediator-mediator interactions  (\cite{nguyen2016causal}, \cite{TchetgenTchetgen2013}, \cite{Wang2013EstimationOC}, \cite{jerolon2021causal}, \cite{nguyen2016causal}). \cite{robins1992identifiability} and \cite{Pearl2001DirectAI} have formulated counterfactual-based definitions of mediation effects.  Hence, we will next define the counterfactual notations in the setting of multiple correlated mediators.

Let $Y(x, M_1(x^*), M_2(x^*))$ denote the counterfactual or potential outcome if $X$ were set to $x$ and both $M_1$ and $M_2$ were set to the values they would take on under $X = x^*$. In the setting of a binary exposure and two mediators, there are eight possible counterfactual outcomes for each individual, not all of which are observed. The causal estimates often of interest in a mediation analysis are the Indirect Effect (IE), and the Direct Effect (DE)  \citep{robins1992identifiability} \citep{Pearl2001DirectAI}. In the context of multiple mediators, the IE is the joint effect of the exposure on the outcome that operates through the group of mediators. Specifically, it assumes that the exposure is set to some level say $X=1$, and then compares the outcome if each mediator had been jointly set to the values they would have taken had the exposure been  $X=1$ versus the values they would have taken had the exposure been $X=0.$
The DE is the effect of the exposure on the outcome that is not mediated by the group of mediators. In many mediation analyses, different scales can be used to express effects. The risk ratio scale is often preferred in epidemiological studies because it provides a direct measure of the probability associated with different levels of exposure. On the risk ratio scale, these effects are defined in terms of the following counterfactuals: 
 $DE = \frac{E(Y(1, M(0,0)))}{E(Y(0, M(0,0)))},$
$IE = \frac{E(Y(1, M(1,1)))}{E(Y(1, M(0,0)))}$
The total effect (TE) is defined as $\frac{E(Y(1, M(1,1)))}{E(Y(0, M(0,0)))}$
and decomposes into the product of the DE and IE on the risk ratio scale. The TE represents the overall causal effect of the exposure on the outcome.

In the multiple mediator setting, path-specific IEs may also be of interest. These effects aim to quantify the contribution of each mediator separately in explaining the exposure-outcome relationship. \cite{Wang2013EstimationOC} noted that there are many different ways to define the path-specific IEs. We used the following definitions of the path-specific IE for mediator 1 (IE1) and for mediator 2
(IE2) respectively: $IE1 =\frac{E(Y(1, M(1,1)))}{E(Y(1, M(0,1)))},$
$ IE2 = \frac{E(Y(1, M(1,1)))}{E(Y(1, M(1,0)))}.$

The path-specific effects defined here can be interpreted as the change in outcome if the expo- sure is held fixed at level $X = 1$ and the mediator of interest changes from the value it would take on under exposure level 0 to the value it would take on under exposure level $1$. In this path-specific effect, the other mediator is set to the value it would take on under exposure level  $X = 1$ in both settings.

To have a causal interpretation of these effects, the four key assumptions for mediation analysis with multiple mediators are:  (1)there is no unmeasured confounding of the exposure-outcome relationship,  (2) no unmeasured confounding of the exposure-outcome relationship (3)no unmeasured confounding of the exposure-mediator relationship, and (4) that the confounders of the mediator-outcome relationship are not affected by the exposure \citep{vanderweele2014mediation}. Additional assumptions are required to identify path-specific effects and will be introduced for the methods designed to identify path-specific effects.

\subsection{Difference-in-coefficients}
The difference-in-coefficients approach, which we call the Difference method throughout this paper, is a straightforward method and is widely utilized in practice. It is one of the traditional methods that was developed by  \cite{BaronKenny1986}. While this method was originally de- signed for normally-distributed variables with linear relationships, it has been used in practice for other distributional scenarios \citep[e.g.][]{douglas2010adverse, howard2018association, carson2021sex}. In a scoping review of mediation analysis methods for observational studies, 57 multiple mediator studies were included, with 8 of these papers using either the Difference method, causal steps, or joint significance approaches to mediation \citep{rijnhart2021mediation}.

The Difference method involves estimating the effect of an exposure on an outcome of interest both with and without the inclusion of the mediators and then comparing these coefficients. This is done by fitting two regression models. The first regression model for the difference method is a regression of Y on X and C:
\begin{equation*}
log[P(Y=1|X, C)] = \phi_0 +\phi_1 X +\boldsymbol{\phi}_2'C
\end{equation*}
The second regression model regresses M, X, and C onto Y:
\begin{equation*}
log[P(Y=1|X, M_1, M_2, C)] = \theta_0 +\theta_1 X +\theta_2 M_1 +\theta_3 M_2 + \theta_4' C
\end{equation*}
 
The coefficient $\phi_1$ represents the TE (or $e^{\phi_1}$ on the RR scale). The DE from X to Y not operating through $M_1$ and $M_2$ is the coefficient $\theta_1$ (or $e^{\theta_1}$ on the RR scale).  The IE, also known as the mediation effect, is the exponentiated difference between the TE and DE in this case: $IE = e^{\phi_1  - \theta_1}$.
 
Here, we used a modified Poisson regression model \citep{zou2004modified} which combines a log-Poisson regression model with robust variance estimation to implement this method. This method estimates just the joint IE but cannot estimate the path-specific IEs. Bootstrapping is often utilized to compute 95\% confidence intervals for the IE and optionally for the other effects.
Due to its ease of implementation, virtually any statistical software package can be used to implement this method.
 
\subsection{VanderWeele and Vansteelandt regression-based}
The causal mediation analysis using the regression-based approach was developed by \cite{valeri2013mediation} and extended to multiple mediators by \cite{vanderweele2014mediation}. Throughout the rest of the paper, this method will be referred to as the Regression method. While regression-based mediation methods were initially proposed by \cite{BaronKenny1986}, this method has been extended to allow for situations where there is an interaction between the exposure and mediator, estimate the causal effects in the counterfactual framework, and allow for non-linear models. This method specifies a separate model for each mediator as well as an outcome model that includes all of the mediators.

In our study, we specify the following modified Poisson model for the outcome assuming the outcome and exposure are binary:
\begin{align*}
log\{P(Y=1|X, M_1, M_2, C)\}  = \theta_0 + \theta_1 X +\theta_2 M_1 +\theta_3 M_2 + \theta_4 C
\end{align*}
The mediator models used depend on the type of mediator variables under consideration. For example, if $M_1$ was a binary mediator and $M_2$ was a continuous mediator, the following two mediator models may be specified:
\begin{equation*}
\begin{aligned}
log\{P(M_1=1|X,C)\} & = \beta_0 +\beta_1 X +\beta_2 C\\
E(M_2|X,C) & = \alpha_0 +\alpha_1 X +\alpha_2 C\\
\end{aligned}
\end{equation*}
\cite{vanderweele2014mediation} used counterfactual-based definitions of each effect to obtain closed-form expressions for the TE, DE, and IE in terms of the model parameters. While closed-form expressions may be derived for certain combinations of mediator and outcome models, they cannot always be obtained. For this reason, an imputation-based version of the regression-based method by \cite{vanderweele2014mediation} is implemented in the \texttt{CMAverse} package in R and is utilized in this study. By using an imputation-based approach, the effects produced can also be interpreted as marginal, rather than conditional causal effects. This method handles cases where there is more than one mediator, regardless to mediator distribution. The outcome variable can be flexible, accommodating various types of data. This method estimates the joint IE but not the path-specific IE effects. It also allows for exposure-mediator interactions but we did not include them in our setting. For more details on this method see \cite{vanderweele2014mediation}. 
 
\subsection{VanderWeele and Vansteelandt weighting-based}
This is a casual mediation analysis method that uses the weighting-based approach by \cite{vanderweele2014mediation}. This approach is based on inverse probability weighting, where the weights are constructed using the exposure and the covariates but not the mediators. This method will be referred to as the Weighting method throughout the rest of the paper. This approach does not require models for the mediators; instead, a model for the exposure is being used. This approach can be used for any type of outcome and exposure but the performance of this method is best when the exposure is binary or discrete with few levels. In this study, we define the exposure and outcome models, respectively, as:
 \begin{equation*}
log[P(X=x|C)] = \phi_0 +\phi_1 C 
\end{equation*}
\begin{equation*}
 log\{P(Y=1|X, M_1, M_2, C)\}  = \theta_0 + \theta_1 X +\theta_2 M_1 +\theta_3 M_2 + \theta_4 C   
\end{equation*}

In order to estimate the natural DE and IEs, this method requires the estimation of three counterfactuals:  $E[Y (1, M(0, 0))]$, $E[Y (1, M(1, 1))]$ and $E[Y (0, M(0, 0))]$. For estimating $E[Y (0, M(0, 0))]$, each subject with $X = 0$ gets a weight of $\frac{P(X_i = 0)}{P(X_i = 0|C_i)}$. $P(X_i = 0|C_i)$ is the predicted probability for each subject with $X = 0$ and can be obtained from a logistic regression if the exposure is binary. The average of each subject with $X = 0$ gives the estimate $E[Y (0, M(0, 0))]$. For $E[Y (1, M(1, 1))]$, each subject with $X = 1$ gets a weight of 
$\frac{P(X = 1)}{P(X = 1|C_i)}$. Again the probability $P(X = 1|C_i)$ can be obtained from a logistic regression if the exposure is binary. To estimate  $E[Y (1, M(0, 0))]$, the outcome model $E[Y=1| X, M_1, M_2, C_i]$ is used to obtain the predicted estimate of the outcome if the individual had had exposure $X_i = 1$ rather than $ X_i = 0$. This estimate is obtained by imputing counterfactuals directly.
After getting the predicted values, $E[Y (1, M(0, 0))]$ is estimated by taking the weighted average of these predicted values. 

On a risk ratio scale, the causal estimates are obtained by taking the ratio of these expectations. Just like the other methods, in multiple mediator settings, this approach estimates the joint IE, but not path-specific IEs. Bootstraps are recommended for constructing confidence intervals. This method is implemented in the \texttt{CMAverse} package \citep{shi2021cmaverse} in R.
\subsection{Inverse odds ratio weighting (IORW)}
Alternative methods for mediation analysis that use a counterfactual-based approach have emerged in recent years, building upon the traditional methods. A popular alternative approach to mediation is the inverse odds ratio weighting (IORW) approach \citep{TchetgenTchetgen2013}.  It is easily implemented with any standard regression model and accommodates multiple mediators of any type.Just like the weighting method, it does not require models for the mediators; instead, a model for the
exposure is being used. This method assigns weight to each observation based on the relationship between the exposure and the mediators, considering other factors like covariates. The weight is the inverse exposure-mediator odds ratio
given covariates which is used to estimate DE via weighted regression analysis. The estimation of the weight differentiate this approach to the weighting method.
The key idea is that this weight helps make the exposure and the mediator act as if they are independent of each other. This is useful because it allows researchers to focus on how the exposure directly affects the outcome when disabling the mediator pathway.
One advantage of IORW is that it examines multiple mediators together, rather than one at a time, by utilizing a model that regresses the set of mediators on the exposure. 

Assuming the exposure is binary, the method can be implemented as follows:
First, fit a logistic regression model on the exposure given the mediators and covariates: \[
\log\left[\frac{P(X=x|M=m,C=c)}{1-P(X=x|M=m,C=c)}\right] = \phi_0 +\phi_1 M_1 + \phi_3 M_2 + \phi_4 C
\]
Next, take the inverse of the predicted odds ratio from the model above for each observation in the exposed group to compute the inverse odds ratio weighting (IORW) weight. The unexposed or control group’s IORW weight will be set to 1. Then, derive the estimated direct effect (DE) by using a modified Poisson model to regress the outcome on the exposure and covariates using the weights:\[
\log[P(Y=1|X=x,C=c)] = \delta_0^* +\delta_1^* X + \delta_3^* C
\]
An estimate of the DE is $e^{\delta_1^*}$. The total effect (TE) can be derived using the same regression model as for the DE but without the weights:\[
\log[P(Y=1|X=x,C=c)] = \delta_0 +\delta_1 X + \delta_3 C
\]
An estimate of the TE is $e^{\delta_1}$. Dividing the DE by the TE will yield the indirect effect (IE) estimate.
The IORW method can also be implemented in any software that incorporates weights in regression analyses and is flexibly implemented in the \texttt{CMAverse} package \citep{shi2021cmaverse} in R. This R package uses an imputation-based version of the IORW method, allowing for marginal mediation effects to be estimated. For more information about this method, see \cite{TchetgenTchetgen2013} and \cite{nguyen2015practical}.

\subsection{Wang et al Integration}
\cite{Wang2013EstimationOC} proposed a counterfactual-based mediation analysis method designed for multiple correlated, contemporaneous mediator variables and a binary outcome. Throughout the rest of the paper, we will refer to this method as Wang's method. This method extends the mediation formula proposed by \cite{pearl2012causal}, which focused on a single mediator, to multiple mediators with possibly mixed types of mediators. This requires a joint mediator distribution for a set of binary and/or continuous mediators to be specified and uses numerical integration techniques to solve the integral in the mediation formula. Bootstrapping is used to obtain 95\% confidence intervals for each of the causal effects of interest. 

This method has an advantage over traditional methods when it comes to multiple mediators. Because the correlation between mediators is captured through the joint mediator model, it is possible to identify path-specific IEs under the additional assumption that the correlation coefficient between each set of mediators is constant and does not change with the value of the exposure \citep{Wang2013EstimationOC}.

This method first requires an outcome model to be specified relating the outcome to the mediators, exposure, and covariate(s):
\begin{equation*}
\begin{aligned}
log(P(Y=1|X,M_1, M_2, C)) = \beta_0 + \beta_1 M_1 + \beta_2 M_2 + \beta_3 X + \beta^ {'}C
\end{aligned}
\end{equation*}
Next, a joint mediator model is specified. In the setting of two continuous mediators, a multivariate normal model is specified, whereas in the case of two binary mediators, a multivariate probit model is specified. When $M_1$ is binary and $M_2$ is continuous, the marginal probit model for the binary mediator and the linear model for the continuous mediator are used:

$$M^{*}_{1} = \alpha_1 X + \alpha_2 C + \epsilon_1$$
$$M_2 =\alpha^{1}_1 X + \alpha^{2}_2 C + \epsilon_2$$
where $M_1$ denotes the binary mediator such that $M^{*}_1$ will is a latent continuous variable underlying $M_1$ denoted as $ M_1 =I\{M_1 \ge 0\}$. The error terms follow a bivariate normal distribution with a mean vector of zeros and a covariance matrix obtained from correlation coefficients between the mediators and standard deviations of the mediators as the variance as shown below: 
      $$ var
      \left (\begin{array}{c}
      \epsilon_1\\
      \epsilon_2\\
      \end{array}\right) =\left(\begin{array}{c c}
      \sigma_1^2 & \rho\sigma_1 \sigma_2\\
      \rho\sigma_1 \sigma_2 & \sigma_2 ^2
      \end{array} \right)
      $$
In the case that $M_1$ is binary, $\sigma_1^2 = 1$, following the typical specification of the probit model.\\    
    Note that we illustrate the estimation approach using two mediators, but this method can be extended to cases with more than two binary and/or continuous mediators. For more details on this method see \cite{Wang2013EstimationOC}. This method was implemented in SAS, with adaptation of the authors' SAS code to compute mediation effects on the risk ratio scale.     

\subsection{J\'{e}rolon  et al quasi-Bayesian}
\cite{jerolon2021causal} developed a mediation analysis method to accommodate mediators that are uncausally related like the method by \cite{Wang2013EstimationOC}. Throughout this paper, we will refer to this method as the J\'{e}rolon's method. While similar to the method by \cite{Wang2013EstimationOC}, the computational strategy differs. This method adopted the approach developed by \cite{imai2010general}, which uses the quasi-Bayesian algorithm to estimate the DE, IE, and IEs through individual mediators using a procedure based on the simulation of counterfactual distributions. This method also proposes using a joint probit and/or multivariate normal mediator model and requires an extra assumption to identify the path-specific effects. For more information about using the algorithm to estimate the causal effects, please see \cite{imai2010general}. This method was implemented in R using the \texttt{multimediate} function from \texttt{multimediate} package. The code from this package was adapted to compute effects on the risk ratio, rather than odds ratio scale.

\subsection{Initial comparison of approaches}
Each of the aforementioned methods brings a unique set of strengths and assumptions to the table; thus, careful consideration of their applicability is essential in the pursuit of accurate and reliable mediation analysis results when multiple mediators are involved. Table \ref{tab:compare} summarizes the functionality and characteristics of each method, as described in Sections 2.1-2.6. 
% \usepackage{graphicx}
\begin{landscape}
\begin{table}[h]
\caption{Comparison of methods for mediation analysis for multiple mediators as implemented currently in available software. RR = relative risk; RD = risk difference; OR = odds ratio.}
\label{tab:compare}
\begin{adjustbox}{max width=\linewidth}
\begin{tabular}{|l|l|l|l|l|l|l|l|l|l|l|l|}
\hline
\textbf{Method}                                                                                                          & \textbf{Abbreviation} & \textbf{IEs}                                      & \textbf{\begin{tabular}[c]{@{}l@{}}Types of \\ Mediators\end{tabular}}   & \textbf{\begin{tabular}[c]{@{}l@{}}Mediator-\\ mediator\\ interaction\\ permitted\end{tabular}} & \textbf{\begin{tabular}[c]{@{}l@{}}Exposure-\\ mediator \\ interaction \\ permitted\end{tabular}} & \textbf{\begin{tabular}[c]{@{}l@{}}Marginal or\\ conditional \\ effect \\ estimates\end{tabular}} & \textbf{\begin{tabular}[c]{@{}l@{}}Contrast for\\ mediation\\ effects\end{tabular}} & \textbf{\begin{tabular}[c]{@{}l@{}}Bootstrapping\\ required\end{tabular}} & \textbf{\begin{tabular}[c]{@{}l@{}}Code\\ availability\end{tabular}}    & \textbf{\begin{tabular}[c]{@{}l@{}}Other\\ considerations\end{tabular}} & \textbf{Computational Burden} \\ \hline
\begin{tabular}[c]{@{}l@{}}Difference-of-coefficients\\ method\end{tabular}                                              & Difference            & Joint                                                          & Any                                                                      & No                                                                                              & No                                                                                                & Conditional                                                                                       & RR                                                                                  & Yes                                                                       & \begin{tabular}[c]{@{}l@{}}R code provided\\ in supplement\end{tabular} & N/A                                                                           & Low                             \\ \hline
\begin{tabular}[c]{@{}l@{}}Inverse odds ratio \\ weighting using \\ imputation in CMAverse\end{tabular}                  & IORW                  & Joint                                                          & Any                                                                      & No                                                                                              & No                                                                                                & Marginal                                                                                          & \begin{tabular}[c]{@{}l@{}}RR, \\ OR\end{tabular}                                   & Yes                                                                       & \begin{tabular}[c]{@{}l@{}}CMAverse\\ package in R\end{tabular}         & N/A                                                                           & Low                             \\ \hline
\begin{tabular}[c]{@{}l@{}}Regression-based approach\\ in CMAverse\end{tabular}                                          & Regression            & Joint                                                          & Any                                                                      & No                                                                                              & Yes                                                                                               & \begin{tabular}[c]{@{}l@{}}Marginal,\\ Conditional\end{tabular}                                   & \begin{tabular}[c]{@{}l@{}}RR, \\ OR\end{tabular}                                   & Yes                                                                       & \begin{tabular}[c]{@{}l@{}}CMAverse\\ package in R\end{tabular}         & N/A                                                                           & Low                             \\ \hline
\begin{tabular}[c]{@{}l@{}}VanderWeele and Vansteelandt\\ weighting approach using\\ imputation in CMAverse\end{tabular} & Weighted              & Joint                                                          & Any                                                                      & Yes                                                                                             & Yes                                                                                               & \begin{tabular}[c]{@{}l@{}}Marginal, \\ Conditional\end{tabular}                                  & \begin{tabular}[c]{@{}l@{}}RR, \\ OR\end{tabular}                                   & Yes                                                                       & \begin{tabular}[c]{@{}l@{}}CMAverse\\ package in R\end{tabular}         & N/A                                                                           & Low                         \\ \hline
Wang et al.                                                                                                              & Wang                  & \begin{tabular}[c]{@{}l@{}}Joint,\\ Path-specific\end{tabular} & \begin{tabular}[c]{@{}l@{}}Continuous,\\ binary\end{tabular}             & No                                                                                              & No                                                                                                & Marginal                                                                                          & RD                                                                                  & Yes                                                                       & \begin{tabular}[c]{@{}l@{}}SAS macro from\\ authors\end{tabular}        & \begin{tabular}[c]{@{}l@{}}Macro written\\ for up to 3 mediators\\ and 1 confounder\end{tabular} & \begin{tabular}[c]{@{}l@{}}Medium for 2 mediators\\ High for more than 2 mediators\end{tabular}                          \\ \hline
J\'{e}rolon et al.                                                                                                           & J\'{e}rolon               & \begin{tabular}[c]{@{}l@{}}Joint,\\ path-specific\end{tabular} & \begin{tabular}[c]{@{}l@{}}Continuous,\\ binary, \\ ordinal\end{tabular} & Yes                                                                                             & Yes                                                                                               & Marginal                                                                                          & \begin{tabular}[c]{@{}l@{}}RD,\\ OR\end{tabular}                                    & No                                                                        & \begin{tabular}[c]{@{}l@{}}Multimediate \\ package in R\end{tabular}    & \begin{tabular}[c]{@{}l@{}}Code accommodates\\ binary confounders only\end{tabular}              & \begin{tabular}[c]{@{}l@{}}Medium\end{tabular}  \\ \hline
\end{tabular}
\end{adjustbox}
\end{table}
\end{landscape}

\section{Simulation study}
In this section, we evaluate the performance of the mediation analysis methods described above. We perform Monte Carlo simulations to compare these methods under different levels of residual correlation, different types of mediators (binary, continuous and mixed), and with or without the interaction effect on the outcome between the mediators. See Table \ref{scenario} for the different scenarios.
\subsection{Simulation set-up}
    To evaluate the existing mediation analysis methods for multiple mediators, we conducted a simulation study involving a binary outcome and two contemporaneous mediator variables. Three main characteristics were varied across simulation scenarios: amount of residual correlation between mediators, type of mediators (two binary, two continuous, or mixed types), and presence or absence of a mediator-mediator interaction. We replicated each scenario using 1000 datasets of size N = 1000. In this simulation study, we examined four metrics for each causal effect (TE, DE, joint IE, and path-specific IEs): percent bias, average 95$\%$ confidence interval width, 95$\%$ confidence interval coverage, and mean squared error (MSE). 200 bootstrap samples were used to construct the 95$\%$ confidence intervals. SAS and R were used for all simulations and statistical analysis.
    
    Across all scenarios, we sampled from the following distributions to generate $X$ and $C$:
    \begin{itemize}
        \item $C \sim Bernoulli (\pi  = 0.5 )$
        \item $X \sim Bernoulli(\pi)$ where    $log(\frac{\pi}{1-\pi}) = -0.25 -C$
        \end{itemize}
     We next created the mediator variables using these samples. In order to achieve this, we generated the error terms for the continuous or underlying continuous random variables from a multivariate normal distribution just as described in the Wang's method.
     We computed the linear predictors ($LP_K$) for each mediator variable as: 
 $$ LP_1 = -1.2 + X + 0.2C + \epsilon_1$$
 $$ LP_2 = -1.5 + 1.5X + 0.5C +\epsilon_2$$
 
 The values of $M_1$ and $M_2$ were then calculated as follows for the data-generating scenarios involving two continuous mediators: $M_1 = LP_1$ and $M_2 = LP_2$. $M_1$ was generated from a probit model as $M_1 = I(LP_1 > 0)$, and $M_2$ was generated from a linear regression model as $M_2 = LP_2$ in the data-generating scenarios with one continuous and one binary mediator. Lastly, we have $M_1 = I(LP_1 > 0)$ and $M_2 = I(LP_2 > 0)$ in the scenarios with two binary mediators. After generating the mediator variables, we generated the binary outcome variable using the model:
 $$ Y \sim Bernoulii(\pi) $$
 $$log(\frac{\pi}{1-\pi}) = \beta_0 + 0.5X +1.5M_1 + 0.5M_2 + \beta_4 M_1M_2 + 1.5C $$
 
 We specifically examined the interaction between $M_1$ and $M_2$ by setting the coefficient $\beta_4$ to 0.2; conversely, for scenarios involving no interaction, we set $\beta_4$ to 0. To capture the non-causal relationships between the mediators, we varied the residual correlation in the mediator values across simulation scenarios, ranging from 0 to 0.75. For every scenario, $\sigma_1$ and $\sigma_2$ were all set to 1. A detailed explanation of every scenario, including the mediators' residual correlation values, is given in Table 2. These effects were selected in order to produce an outcome proportion that reflects real-world scenarios, ranging from 0.25 to 0.4. Since there is no closed-form solution for these effects, we used Pearl's mediation formula along with the \texttt{adaptintegrate} function from the \texttt{cubature} package \citep{hahn2005cuba} in R to closely approximate the true causal effects. 
\begin{table}[ht]
\caption{Scenarios description for the simulation study where interaction indicates the presence or absence of an interaction between mediators and residual correlation describes the correlation of the error terms in the mediator models.}
\label{scenario}
\centering
\begin{tabular}{|c|c|c|c|}
\hline
\textbf{Scenario} & \textbf{Mediator Type} & \textbf{Residual Correlation} & \textbf{Interaction} \\
\hline
1 & 2 Continuous & 0 & No \\
2 & 2 Binary & 0 & No \\
3 & 1 Binary, 1 Continuous & 0 & No \\
4 & 2 Continuous & 0.25 & No \\
5 & 2 Binary & 0.25 & No \\
6 & 1 Binary, 1 Continuous & 0.25 & No \\
7 & 2 Continuous & 0.5 & No \\
8 & 2 Binary & 0.5 & No \\
9 & 1 Binary, 1 Continuous & 0.5 & No \\
10 & 2 Continuous & 0.75 & No \\
11 & 2 Binary & 0.75 & No \\
12 & 1 Binary, 1 Continuous & 0.75 & No \\
13 & 2 Continuous & 0 & Yes \\
14 & 2 Continuous & 0.25 & Yes \\
15 & 2 Continuous & 0.5 & Yes \\
16 & 2 Continuous & 0.75 & Yes \\
\hline
\end{tabular}
\end{table}

\subsection{Simulation results}
\begin{figure}
\centering
\includegraphics[width=0.9\textwidth]{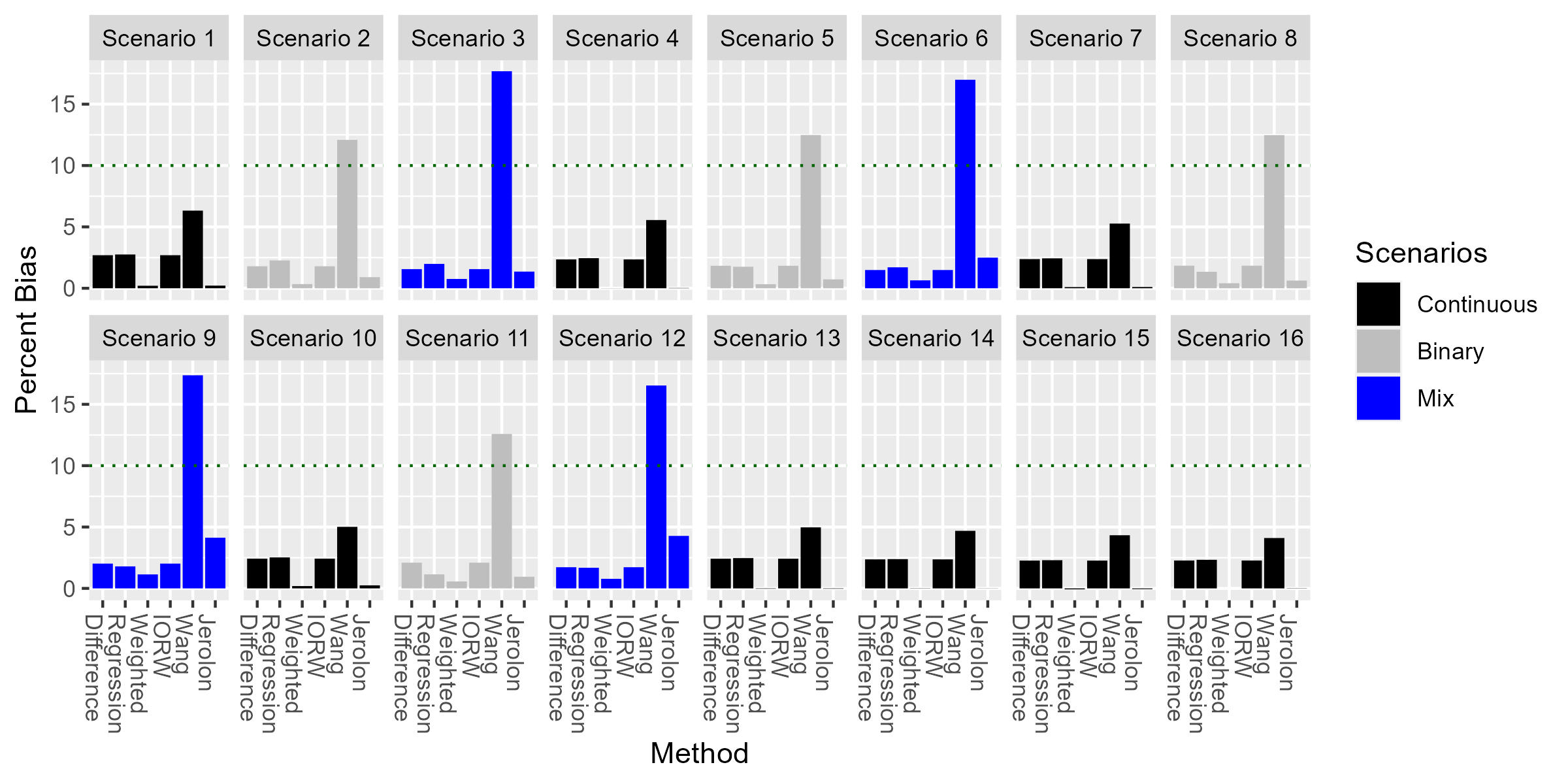}
\caption{Percent Bias for TE for each method across all scenarios. The scenarios are arranged such that residual correlation increases from left to right: scenarios 1-3 have no residual correlation, scenarios 4-6 have a residual correlation of 0.25, and so on. The last four scenarios depict those with mediator-mediator interactions, with scenario 13 illustrating a setting with no residual correlation and scenario 16 illustrating a setting with a residual correlation level of 0.75.}
\label{Figure:A}
\end{figure}

\begin{figure}
\centering
\includegraphics[width=0.9\textwidth]{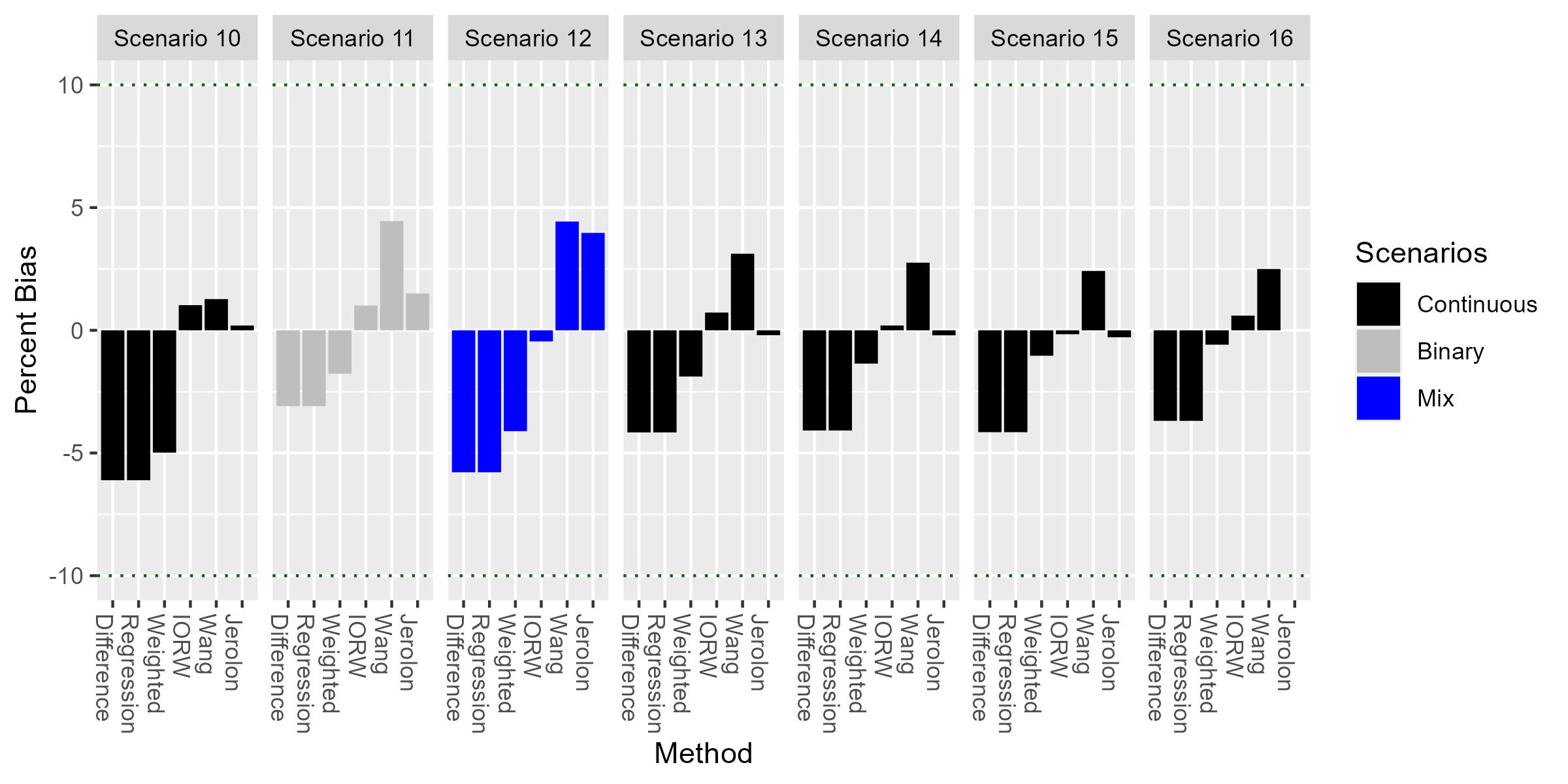}
\caption{Percent Bias for DE across methods in scenarios 10-16. Scenarios 10-12 show differing mediator types with a residual correlation of 0.75. Scenarios 13-16 include interactions between mediators, with increasing residual correlation from 13 to 16.}
\label{Figure:B}
\end{figure}

\begin{figure}[ht]
\centering
\includegraphics[width=0.9\textwidth]{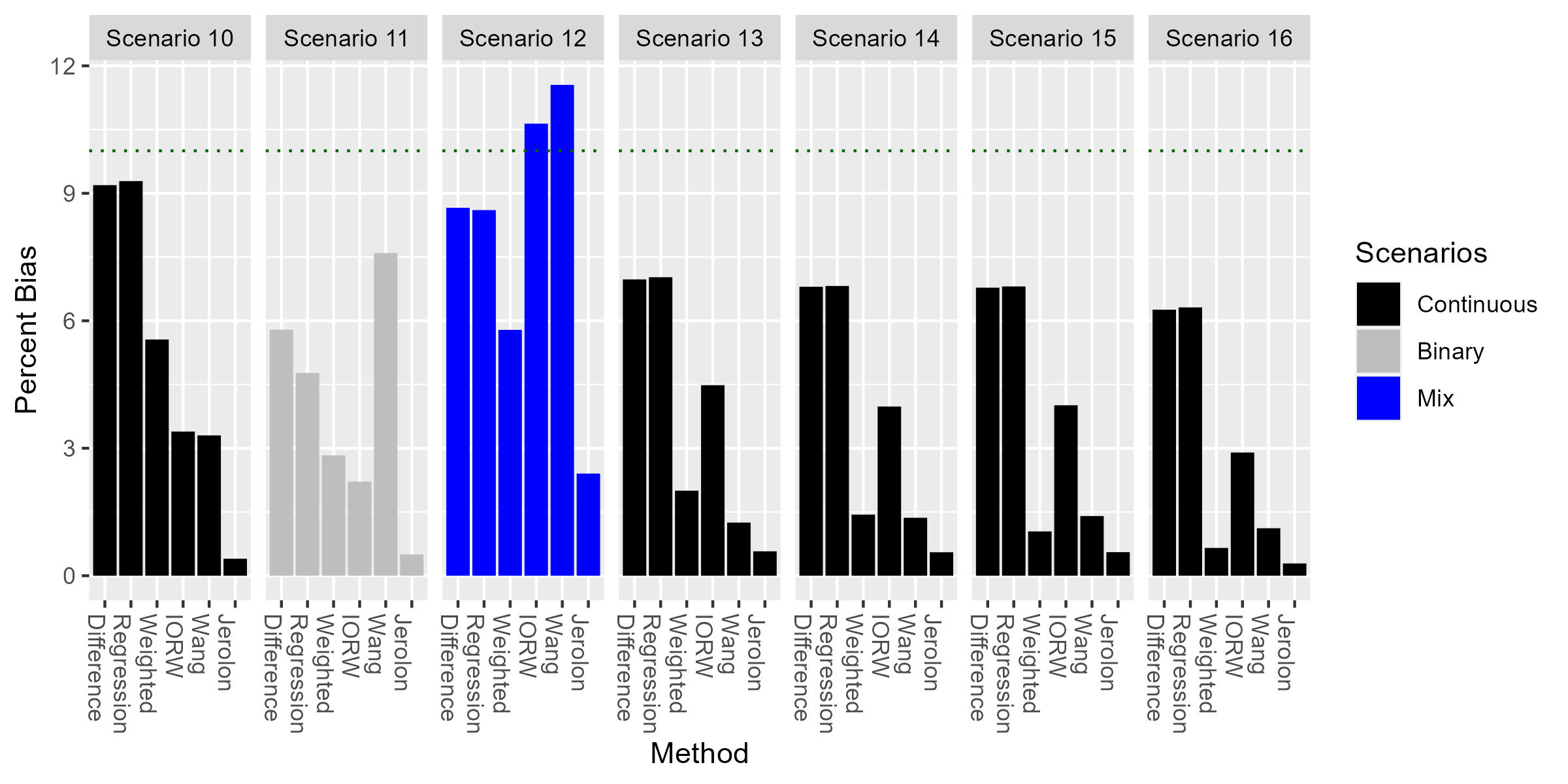}
\caption{Percent Bias for IE across methods in scenarios 10-16. Scenarios 10-12 show differing mediator types with a residual correlation of 0.75. Scenarios 13-16 include interactions between mediators, with increasing residual correlation from 13 to 16.}
\label{Figure:C}
\end{figure}

Figures \ref{Figure:A}, \ref{Figure:B}, and \ref{Figure:C} illustrate the percent bias for the TE, DE, and IE, respectively, across all scenarios. In each figure illustrating the percent bias, a dotted line is included at 10$\%$ and at -10 $\%$ to indicate bias in the negative direction. For coverage, a dotted line is shown at 0.95. These lines are provided for consistent comparison across figures.  Note that in each set of figures, residual correlation increases from left to right (e.g. scenarios 1-3 have no residual correlation, 4-6 have a residual correlation of 0.25, etc.). The last four scenarios depict those with interaction effects between mediators, where scenario 13 illustrates a setting with no residual correlation and scenario 16 illustrates a setting with a residual correlation level of 0.75. See Table \ref{scenario} for a complete description of each scenario.

In terms of the TE, the weighted method consistently exhibited the least bias across all scenarios, followed by the method developed by J\'{e}rolon. Wang’s method displayed higher bias compared to other methods for the TE, typically exceeding 10$\%$. Particularly notable was its elevated bias when analyzing scenarios featuring a mixture of binary and continuous mediators. However, in scenarios where all mediators were continuous, Wang’s method demonstrated comparatively lower percent bias, albeit still higher than other methods. Interestingly, as the residual correlation between mediators increased, the percent bias in the TE for Wang's method reduced for continuous mediators, a trend observed consistently across scenarios with or without a mediator-mediator interaction.
 Furthermore, the IORW and Difference methods demonstrated similar percent bias for the TE across all scenarios, with the Regression method performing comparably in scenarios involving two continuous mediators, with or without interactions. Notably, the Regression method exhibited lower bias compared to IORW and the Difference method for the TE in scenarios involving all binary mediators with a residual correlation greater than or equal to 0.5 (scenarios 8 and 11).

When analyzing the percent bias in the DE (Figure \ref{Figure:B}), percent bias occurred in both positive and negative directions. Scenarios without residual correlation and with increasing residual correlation show similar pattern for continuous, binary and mix mediator types and hence we show scenarios with residual correlation of 0.75 as well as scenarios with interaction. The Difference, Regression, and Weighted methods tended to exhibit bias in the negative direction, while IORW, Wang, and J\'{e}rolon methods showed bias in the positive direction. J\'{e}rolon performed favorably for scenarios involving two continuous mediators, followed by Wang’s method. Conversely, IORW demonstrated better performance in scenarios with two binary mediators and also did quite well estimating the DE in Scenarios 13-16, which had continuous mediators with a mediator-mediator interaction. In all continuous and mixed mediator scenarios, the Difference and Regression methods exhibited percent biases hovering around $5\%$ or more in the negative direction and did not perform relatively well compared to the other methods. However,their percent bias were not very huge. 

Examining Figure \ref{Figure:C}, we observe that J\'{e}rolon’s method consistently displayed the lowest percent bias for the IE across all scenarios, whereas Wang's method exhibited the highest bias in mix and binary scenarios. Across all scenarios, the Difference and Regression methods exhibited consistently high percent biases for the IE. However, there was no single method consistently exhibiting the highest percent bias across all scenarios; rather, it varied depending on the type of mediator variables involved. For instance, in scenarios with two continuous mediators, the Difference and Regression methods tended to have the highest percent bias. Additionally, the percent bias for all methods tended to decrease marginally as residual correlation increased, particularly evident in scenarios where mediators are continuous. With respect to mediator-mediator interactions, Wang's method performed quite well in Scenarios 13-16 in which a mediator-mediator interaction was present, despite this method not explicitly an interaction effect. In addition, the Weighted and J\'{e}rolon methods did quite well in the presence of a mediator-mediator interaction. This was less surprising, as they were the two methods which allowed for a mediator-mediator interactions to be specified in the outcome model.

For each effect, we assessed coverage of the 95\% confidence intervals. For the TE, coverage levels typically hovered between 90\% and 95\%, with the exception of Wang's method, which had coverage levels near 75\% for the TE in almost all scenarios. Coverage levels were quite similar for the DE across all of the scenarios involving binary or mixed mediators but differed in the two continuous mediator scenarios. Notably, scenarios involving all continuous mediators revealed that two methods demonstrated improved coverage over the others for the DE: the J\'{e}rolon and Wang methods. In contrast, the Regression and Difference methods always had unsatisfactory coverage for the DE in the continuous mediator settings.  
\begin{figure}
\centering
\includegraphics[width=0.9\textwidth]{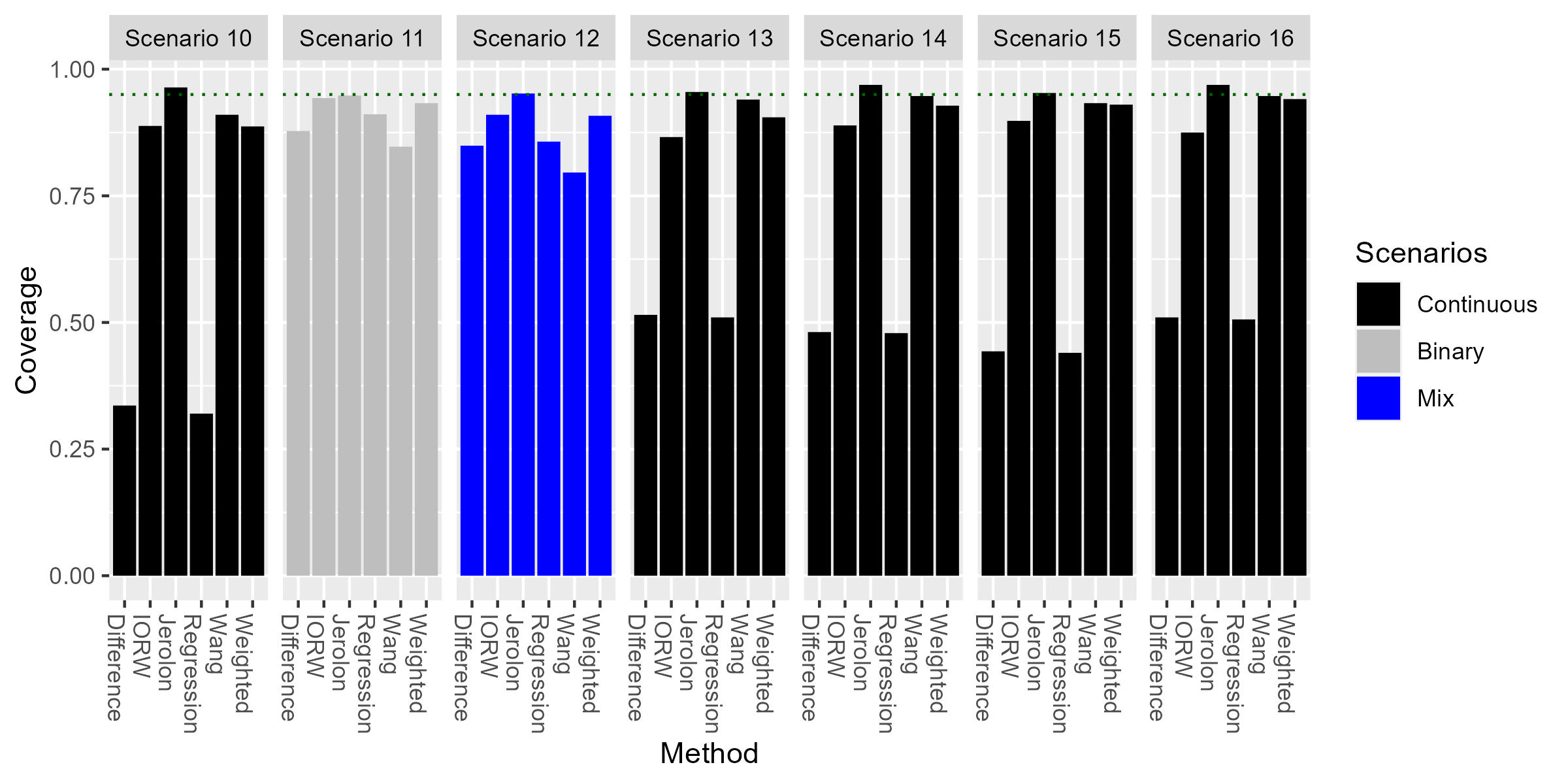}
\caption{Coverage of IE across methods in scenarios 10-16. Scenarios 10-12 show differing mediator types with a residual correlation of 0.75. Scenarios 13-16 include interactions between mediators, with increasing residual correlation from 13 to 16.}
\label{Figure:F}
\end{figure}

Figure \ref{Figure:F} displays the coverage of the 95\% confidence intervals for the IE across scenarios. A similar trend is observed to the coverage for the DE confidence intervals; however, the coverage levels are generally lower. J\'{e}rolon's method again demonstrated the highest coverage compared to the other methods. Notably, in continuous mediator settings, the Difference, IORW, and Regression methods exhibited very suboptimal performance, with coverage as low as 50$\%$ in some scenarios and even dropping below 30$\%$  in certain continuous settings. 

The results for average 95\% confidence interval width suggested two key patterns. For the TE, methods typically performed similarly except for Wang's method, which produced wider intervals in the mixed mediator setting. For the DE, the IORW method stood out as having intervals that were almost twice the width of the others, and this pattern persisted for the IE confidence intervals. Further details regarding average interval width are available in the supplementary figures. Because RMSE represents a tradeoff between bias and variance, both of which have been discussed, we leave these results for the supplementary materials as well.

\begin{figure}[ht]
\centering
\includegraphics[width=0.9\textwidth]{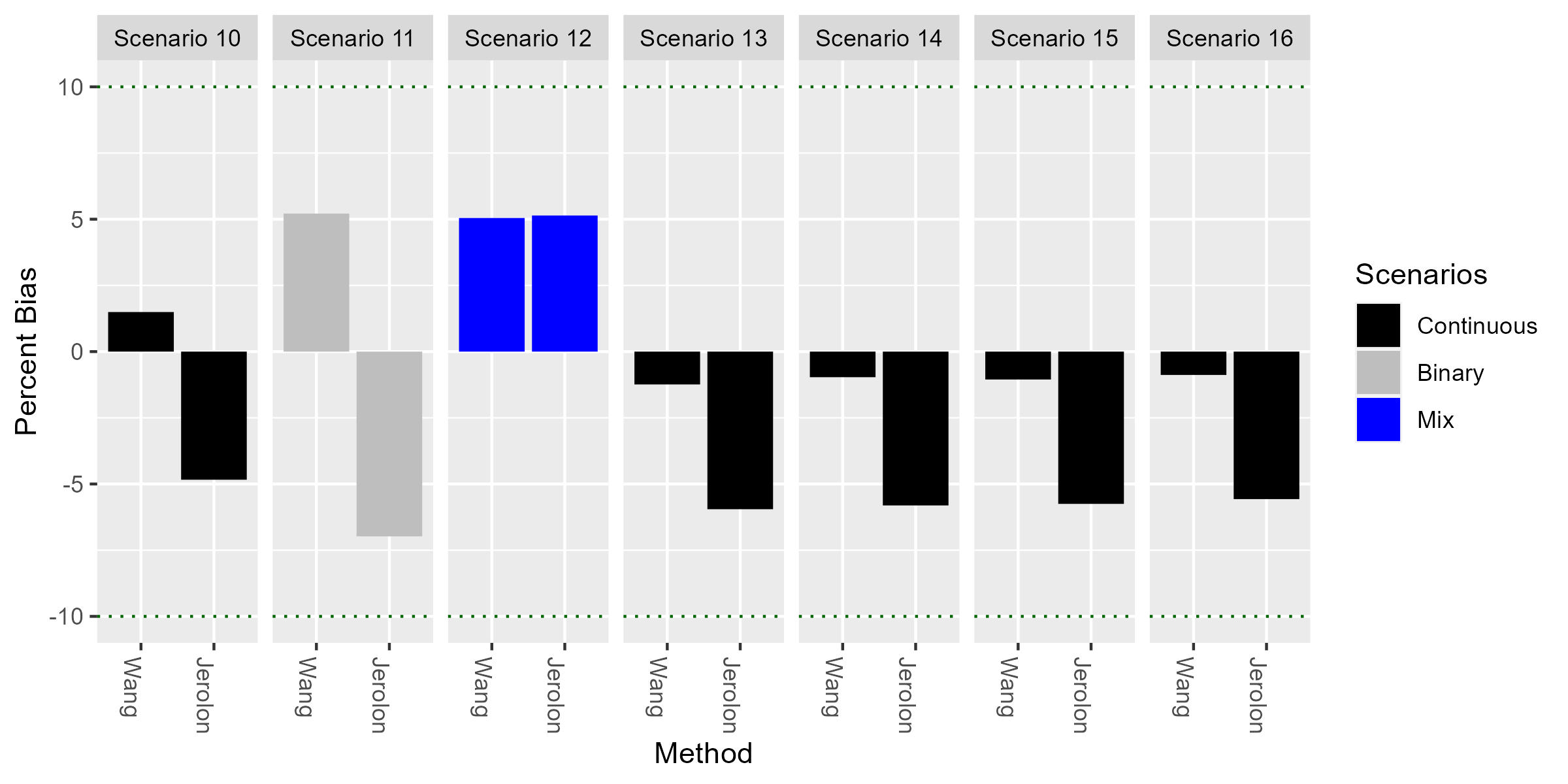}
\caption{Percent Bias for IE1 across methods in scenarios 10-16. Scenarios 10-12 show differing mediator types with a residual correlation of 0.75. Scenarios 13-16 include interactions between mediators, with increasing residual correlation from 13 to 16.}
\label{Figure:G}
\end{figure}

\begin{figure}[ht]
\centering
\includegraphics[width=0.9\textwidth]{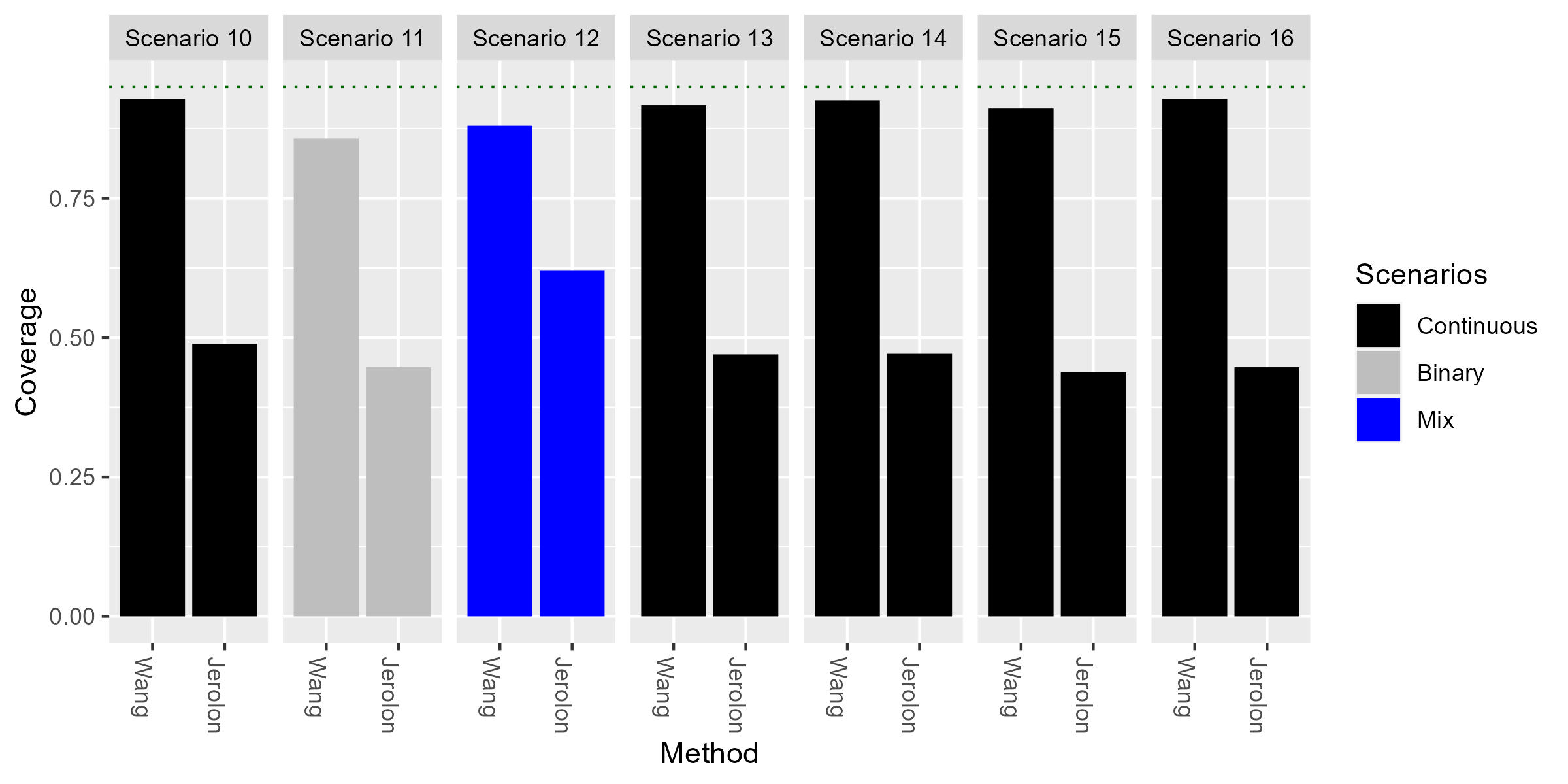}
\caption{Coverage of IE1 across methods in scenarios 10-16. Scenarios 10-12 show differing mediator types with a residual correlation of 0.75. Scenarios 13-16 include interactions between mediators, with increasing residual correlation from 13 to 16.}
\label{Figure:I}
\end{figure}

Among the six methods compared, two were capable of calculating path-specific IEs. For the percent bias of IE1 (Figure \ref{Figure:G}), Wang's method outperformed J\'{e}rolon's method, particularly in scenarios with continuous mediators, albeit with a lower magnitude. Notably, the direction of the bias varied across continuous and binary mediator types but remained consistent for mixture mediator types. However, for interactions involving continuous mediators, the bias direction was consistent but negative. Regarding coverage (Figure \ref{Figure:I}), Wang's method consistently exhibited superior performance with coverage, whereas J\'{e}rolon's method fell far below the 95\% threshold. Wang's method also generally produced wider confidence intervals for the path-specific effects than J\'{e}rolon's method, with the exception of the continuous mediator scenarios. \cite{jerolon2021causal} also indicated that their method produced suboptimal coverage for path-specific effects in the binary outcome setting, and our results agree with these findings. Regarding the percent bias for the IE2, J\'{e}rolon's method performed better than Wang's method, albeit in the opposite direction (positive for Wang and negative for J\'{e}rolon) in scenarios involving one binary and one continuous mediator. However, Wang's method outperformed J\'{e}rolon's method across both binary and continuous mediator settings, albeit with a lower magnitude in continuous mediator settings. In most scenarios, the Wang's method more accurately estimated the path-specific IEs with an improved coverage.

\section{Application of methods to understand socioeconomic differences in hypertension incidence}
We applied four of the mediation analysis methods (Difference, Regression, IORW, and Weighted) to the REasons for Geographic And Racial Differences in Stroke (REGARDS) study \citep{howard2005reasons} to understand mechanisms explaining socioeconomic differences in incident hypertension. These four methods were selected for application due to their fast computation time with large datasets, incorporate more than three mediators and their ability to examine the collective effect of a larger set of mediator variables, which has been a goal of many recent analyses performed by REGARDS investigators \citep[e.g.][]{howard2018association, tajeu2020black, carson2021sex}. Use of these methods allowed us to examine the TE, DE, and IE but not the path-specific IEs.

REGARDS is a large national cohort study with more than 30,000 participants. All participants are Black and White adults residing in the United States who were at least 45 years old at the initial visit. Data were collected at baseline through an in-home visit, telephone interview, and self-administered questionnaires. The data collected included physical and physiological measurements, demographic characteristics, dietary patterns, lifestyle information, a medication history, and more. A second follow-up visit occurred approximately ten years later, where physical and physiological measurements were taken again.

In this application, our exposure was low household income at baseline, defined as a household income of $< \$$35,000 per year. Participants who did not report their household income were excluded from the analysis. Our outcome was incident hypertension, defined as a new hypertension case by the second in-home visit. Specifically, we used the JNC7 definition of hypertension (systolic blood pressure $\geq$ 140, diastolic blood pressure $\geq$ 90, or self-reported current medication use to control blood pressure) \citep{chobanian2003septieme} and excluded participants with hypertension at baseline. The four mediators under consideration were current smoking status (binary - current smoker vs. not a current smoker), any exercise (binary - reported exercising 1+ times per week vs. 0 times per week), Dietary Approaches to Stop Hypertension (DASH) diet score (continuous), and Dietary Inflammation Score (DIS) (continuous). The DASH diet has been shown to lower blood pressure and is rich in nutrients including potassium and manganese. Higher values indicate greater adherence to the DASH diet \citep{appel2006dietary}. DIS is a dietary score comprised of primarily whole foods, where higher values reflect more pro-inflammatory dietary exposures \citep{byrd2019development}. All mediation analyses adjusted for systolic blood pressure, education level, race, sex, age (included as a quadratic term), and region (stroke belt, stroke buckle, or other) at baseline, and participants with complete data on the aforementioned variables were included in this analysis.

In total, 5,346 participants met the criteria for inclusion. Of these participants, 45.0\% were male and 20.8\% were Black. Incident hypertension occurred in 35.3\% of the participants, and 27.0\% of participants had a household income of $<$ \$ 35,000 at baseline. Of those in the low-income group, 57.6\% had incident hypertension, compared to 32.7\% in the higher-income group. The two dietary scores, DIS and DASH, had a moderate negative correlation of -0.634.

Table \ref{tab:regards} presents the effect estimates, 95\% confidence intervals, and interval widths for the TE, DE, and IE under the Difference, Regression, Weighted, and IORW methods. Interestingly, the estimated TE was similar among the Difference, Regression, and IORW methods at about 1.15, whereas the Weighted method produced a TE estimate of 1.12. This aligns with the simulation results in Figure \ref{Figure:A}, in which the percent bias was similar for the Difference, Regression, and IORW methods in the mixed mediator scenarios but lower for the Weighted Method. In terms of the DE, the Difference, Regression, and Weighted methods produced similar estimates of 1.13, whereas the IORW method produced a DE estimate of 1.11. The 95\% confidence interval for the DE excluded 1 under the Difference, Regression, and Weighted methods but included 1 under the IORW method. 

In terms of the IE, which is often the effect of primary interest in a mediation analysis, the estimates were greater than 1 under the Difference, Regression, and IORW methods but was estimated to be 0.99 under the Weighted method. In both weighting-based methods (Weighted and IORW), the 95\% confidence intervals for the IE included 1. In contrast, the confidence intervals constructed using the Difference and Regression methods excluded 1, allowing us to conclude that the collective effect of smoking, physical activity, DIS, and DASH dietary score explain some of the difference in hypertension by socioeconomic status. The weighting-based methods also produced wider intervals than the Difference and Regression methods, consistent with the simulation results. However, as illustrated in the simulation, the Difference and Regression methods tended to produce a biased decomposition of the TE into the DE and IE. The results of this analysis illustrate that the four mediation analysis methods produce different results and conclusions in real data analyses in the presence of correlated mediator variables of different types (continuous and binary). These findings, combined with our simulation results, highlight the importance of considering which mediation analysis method best fits the research question and dataset.


\begin{table}[h]
\centering
\caption{Effect estimates, 95\% confidence intervals, and width of the confidence interval for the TE, DE, and IE when applying the Difference, Regression, Weighted, and IORW methods to the REGARDS dataset. All estimates are on the RR scale.}
\label{tab:regards}
\centering
\begin{tabular}{|l|p{4cm}|p{4cm}|p{4cm}|}
\toprule
\textbf{Approach} & \textbf{Total Effect} & \textbf{Direct Effect} & \textbf{Indirect Effect} \\ \midrule
\textbf{Difference} & 1.152 \newline (1.064 – 1.251) \newline 0.187 & 1.129 \newline (1.041 – 1.226) \newline 0.185 & 1.021 \newline (1.009 – 1.035) \newline 0.026 \\ \midrule
\textbf{Regression} & 1.151 \newline (1.063 – 1.252) \newline 0.189 & 1.129 \newline (1.041 – 1.226) \newline 0.185 & 1.020 \newline (1.009 – 1.034) \newline 0.025 \\ \midrule
\textbf{Weighted} & 1.121 \newline (1.021 – 1.225) \newline 0.204 & 1.129 \newline (1.043 – 1.222) \newline 0.179 & 0.994 \newline (0.949 – 1.037) \newline 0.088 \\ \midrule
\textbf{IORW} & 1.152 \newline (1.065 – 1.251) \newline 0.187 & 1.111 \newline (0.999 – 1.231) \newline 0.232 & 1.036 \newline (0.973 - 1.106) \newline 0.133 \\ \bottomrule
\end{tabular}
\end{table}

\section{Discussion}
In this paper, we assessed the performance of various mediation analysis methods under diverse scenarios, including varying levels of residual correlation among mediators, different mediator types, and the presence of interaction effects between mediators. The findings from our simulation study and data application provide valuable insights into the strengths and limitations of each method across different conditions.

The methods that tended to perform best across various scenarios were the Weighted, IORW, J\'{e}rolon, and Wang methods. The Weighted method consistently exhibited low bias in estimating TEs, despite the data-generating mechanism aligning more closely with J\'{e}rolon and Wang's methods. Its robustness in accurately estimating TEs, ability to handle interaction effects, and relatively low computation time are noteworthy strengths. However, limitations include its inability to estimate path-specific effects and occasionally having wider CIs. The J\'{e}rolon method provided estimates of TEs, DEs, and IEs with low bias. It uses a quasi-Bayesian algorithm for estimation which is faster compared to  numerical integration methods and bootstrapping employed by Wang's method. Additionally, its implementation in an R package enhances accessibility. However, limitations include a tendency for high bias and low coverage of path-specific effects, and it can currently incorporate only binary confounders. This was another challenge that precluded its use in the data application. 
The IORW method, known for its flexibility in allowing any type of mediator variables and ease of implementation in standard regression software capable of handling weights, demonstrated relatively good performance in certain scenarios. It exhibited very low bias in estimating DE in binary and continuous mediator settings and performed well with mediator interactions, particularly in scenarios with moderate residual correlation. However, similar to the Weighted method, it cannot estimate path-specific effects and may result in wider confidence intervals, especially in mixed settings.
The Wang's method stands out for its ability to estimate path-specific effects with low bias and higher coverage. However, it has limitations including high computational time, difficulty in implementation, and wider confidence intervals for TE in mixed scenarios. In general, it doesn't perform well in mixed mediator settings.

In our setting, we focused on a binary outcome because it is very common in epidemiological studies. Causal mediation analysis methods are preferred when performing mediation analysis with nonlinear regression. The Difference and product-of-coefficient methods are usually referred to as ``traditional'' methods \citep{rijnhart2021mediation}. Effect estimation in the Difference method is most closely related to the Regression method  \citep{vanderweele2014mediation}, which estimates the IE using the product-of-coefficient method in the absence of exposure-mediator interaction \citep{breen2013total,rijnhart2023statistical,pearl2012causal}. We observed this in both simulation and data application results, that is, similarities in results  between the Difference and
Regression method. These methods were observed to perform poorly in almost all scenarios. It has been established in \cite{richiardi2013mediation} that these traditional methods estimate biased effects in the presence of unmeasured confounders. To address this issue, \cite{richiardi2013mediation} recommended using the mediation formula which has been used in the Wang's method and the inverse probability weighted which is used by \cite{vanderweele2014mediation} which we refer to in our paper as the Weighted method. The traditional methods are still being used and should not be discarded, but researchers should understand when to use them. Furthermore, we noted that despite the similarities in their model formulations, Wang and J\'{e}rolon's methods yielded markedly different results. While both methods have the same model formulations, their approaches diverge in estimating causal effects. Wang's method uses the parameters for estimation, on the other hand, J\'{e}rolon's approach uses the predicted y’s during estimation, potentially rendering it more robust to model misspecification.

Interestingly, the mediator type seems to be more influential in the choice of method than the amount of correlation between mediators. The influence of mediator type on method choice is a noteworthy observation stemming from our comparison of methods. While one might expect the correlation between mediators to have a significant impact on method performance, our findings suggest that the type of mediator variable exerts a stronger influence. This finding underscores the importance of carefully considering the nature of the mediators when selecting an appropriate causal mediation analysis method. For instance, in scenarios involving binary mediators, certain methods such as the J\'{e}rolon's method exhibited favorable performance, whereas in scenarios with continuous mediators, different methods such as the Wang's method performed better. 

While our study provides valuable insights into the comparison of different mediation analysis methods, it is crucial to acknowledge certain limitations that could impact the interpretation and practicality of our findings. One significant consideration is the distributional assumptions we made in our simulations study. We assumed that continuous mediators followed a normal distribution and did not explore heavy-tailed or skewed continuous distributions. Additionally, our simulation study involved binary outcomes, exposures, and a binary baseline covariate which inherently represent a limited spectrum of scenarios compared to the complexities observed in real-life situations. However, we also explored four of the methods' performances in a data example with multiple baseline covariates of mixed types and arrived at similar conclusions in a real-world scenario. To address these limitations, future simulation studies could delve into the distributional characteristics of the variables under examination. Employing transformation methods to normalize skewed distributions before comparing mediation analysis techniques may also enhance the robustness of our findings. Secondly, our study focused on contemporaneous mediators and did not consider the inclusion of mediators with an inherent ordering. Future work is needed to examine the performance of the methods included in this study, along with methods specifically designed for decomposing IEs with ordered mediators \citep[e.g.][]{daniel2015causal, gao2019bayesian} in these settings.

\section{Conclusion}
\label{sec:conc}
In conclusion, our study underscores the importance of methodological considerations and empirical validation in mediation analysis. Our study aims to equip practitioners with insight and guidance into the best mediation analysis methods for their multiple mediator study. By selecting a mediation method that is targeted to their scientific question of interest, practitioners can further contribute to the advancement of knowledge in their respective fields.

\pagebreak
\begin{center}
{\large\bf SUPPLEMENTARY MATERIAL}
\end{center}
\begin{figure}[H]
\begin{center}
\includegraphics[width=0.7\textwidth]{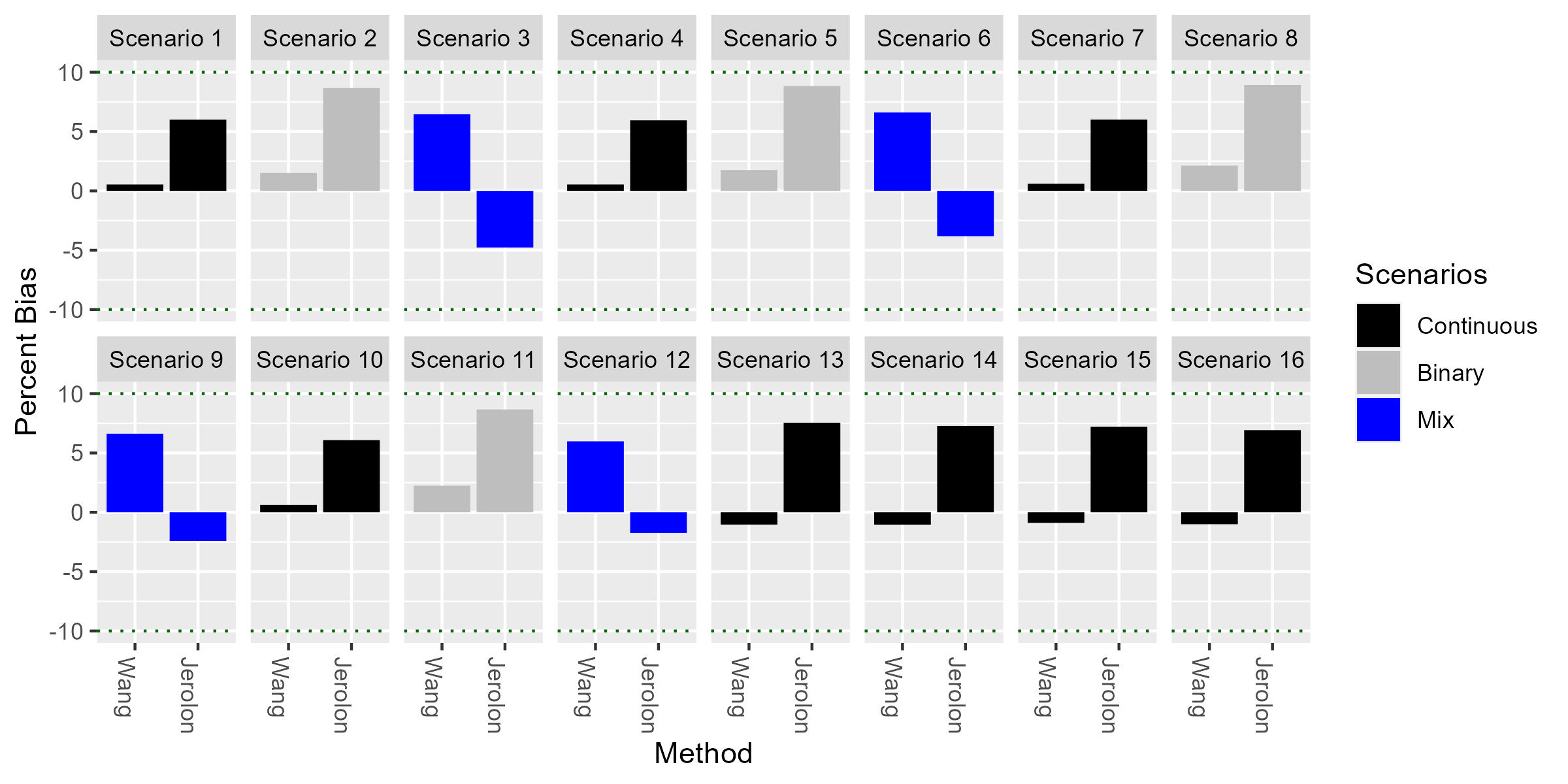}
\end{center}
\caption{Percent Bias of IE2 across various scenarios as shown in \ref{scenario} The scenarios are arranged such that residual correlation increases from left to right: scenarios 1-3 have no residual correlation, scenarios 4-6 have a residual correlation of 0.25, and so on. The last four scenarios depict those with interaction effects between mediators, with scenario 13 illustrating a setting with no residual correlation and scenario 16 illustrating a setting with a residual correlation level of 0.75\label{Figure:H}}
\end{figure}

\begin{figure}[H]
\begin{center}
\includegraphics[width=0.7\textwidth]{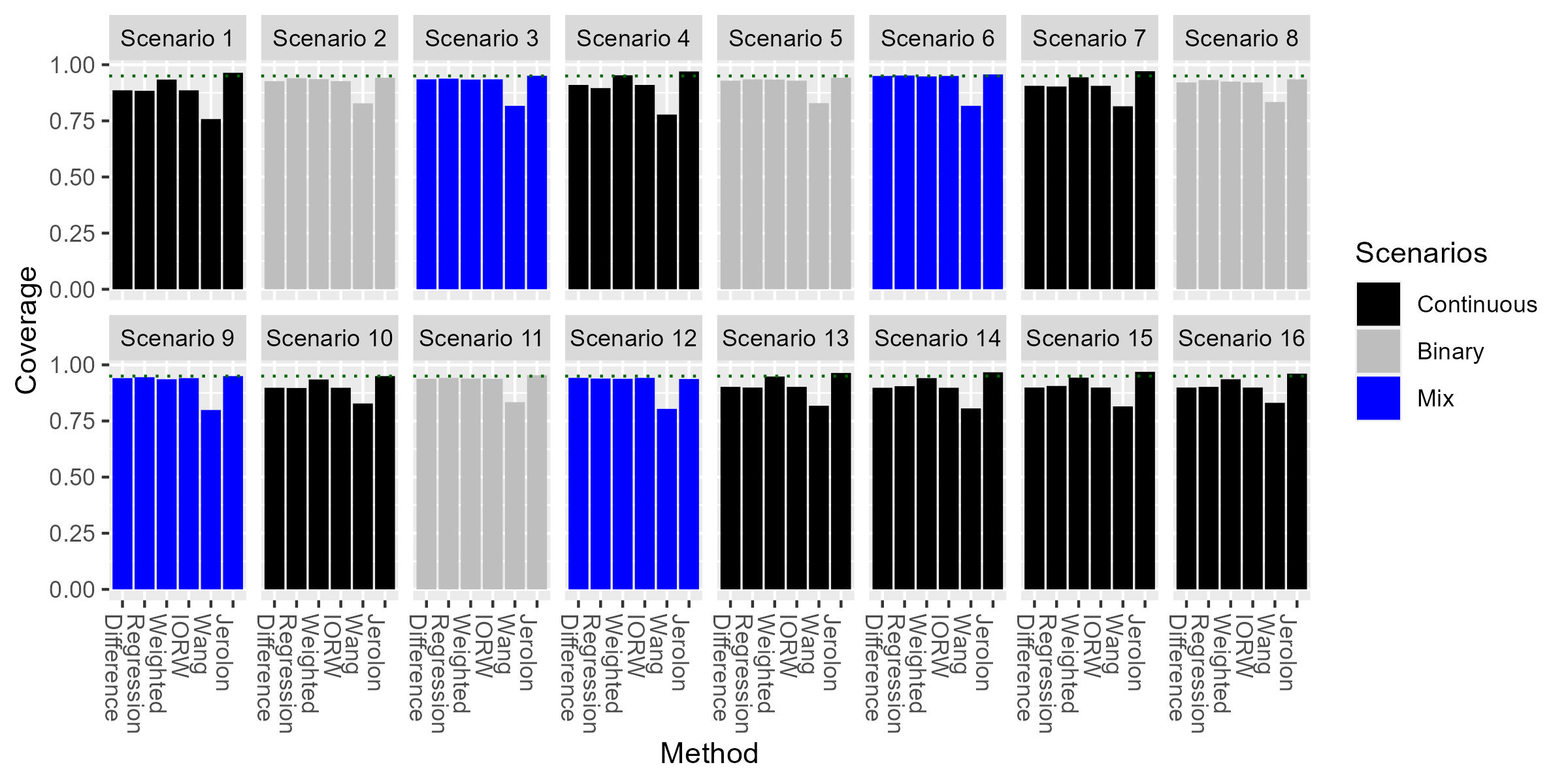}
\end{center}
\caption{Coverage of TE across various scenarios as shown in \ref{scenario} The scenarios are arranged such that residual correlation increases from left to right: scenarios 1-3 have no residual correlation, scenarios 4-6 have a residual correlation of 0.25, and so on. The last four scenarios depict those with interaction effects between mediators, with scenario 13 illustrating a setting with no residual correlation and scenario 16 illustrating a setting with a residual correlation level of 0.75\label{Figure:D}}
\label{Figure:J}
\end{figure}

\begin{figure}[H]
\begin{center}
\includegraphics[width=0.7\textwidth]{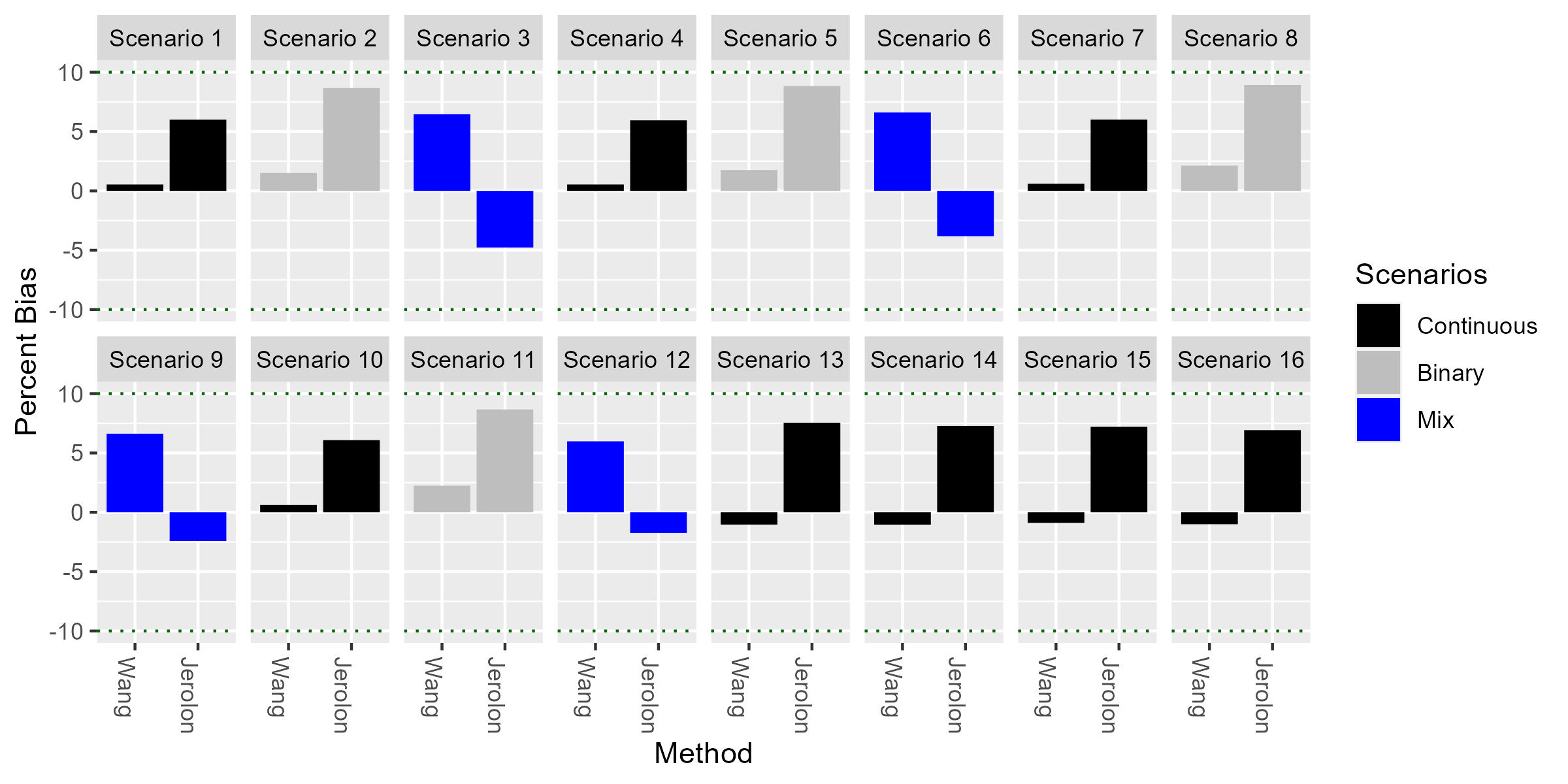}
\end{center}
\caption{Coverage of IE2 across various scenarios as shown in \ref{scenario} The scenarios are arranged such that residual correlation increases from left to right: scenarios 1-3 have no residual correlation, scenarios 4-6 have a residual correlation of 0.25, and so on. The last four scenarios depict those with interaction effects between mediators, with scenario 13 illustrating a setting with no residual correlation and scenario 16 illustrating a setting with a residual correlation level of 0.75}
\end{figure}

\begin{figure}[H]
\begin{center}
\includegraphics[width=0.7\textwidth]{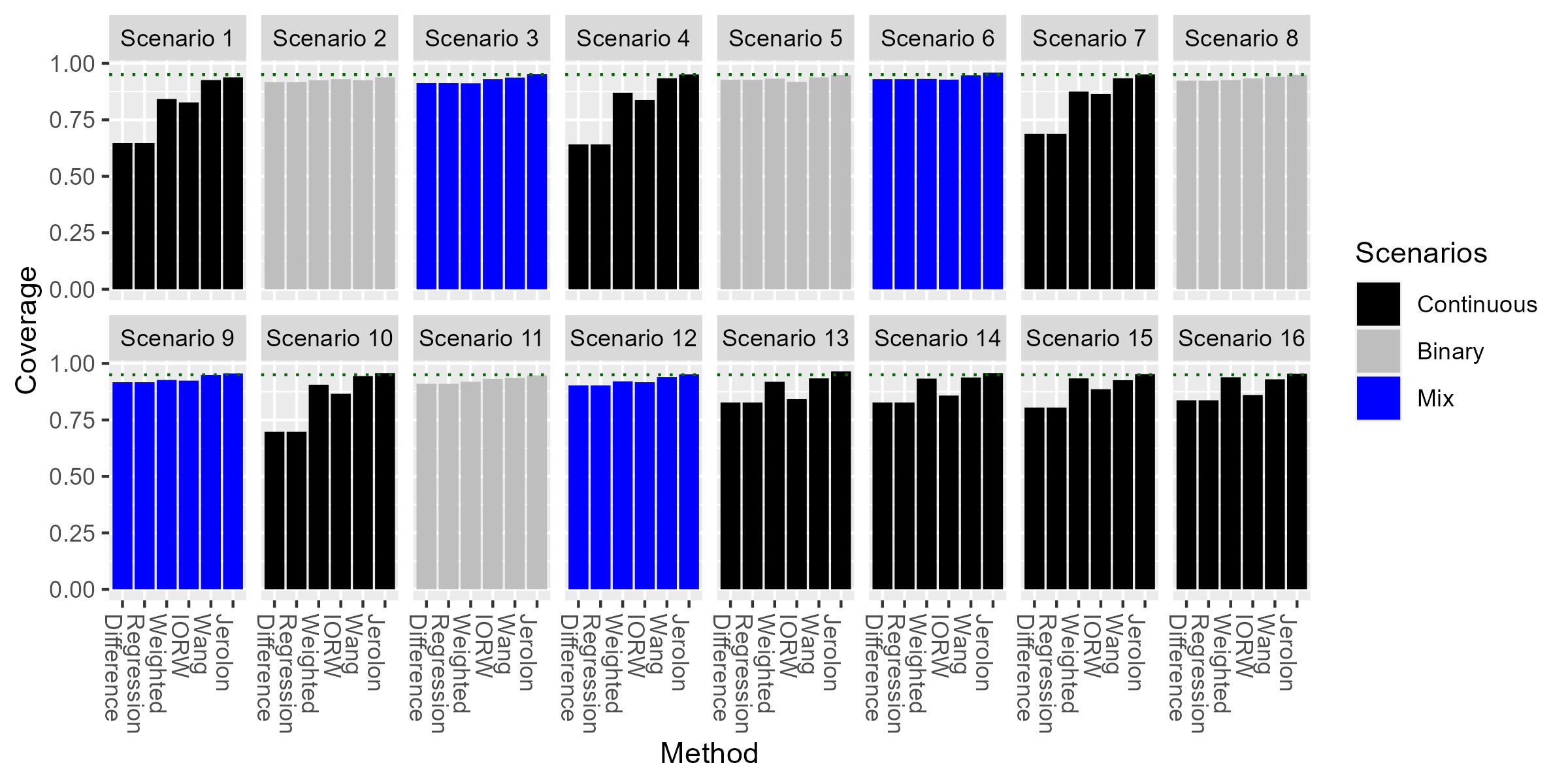}
\end{center}
\caption{Coverage of DE across various scenarios as shown in \ref{scenario} The scenarios are arranged such that residual correlation increases from left to right: scenarios 1-3 have no residual correlation, scenarios 4-6 have a residual correlation of 0.25, and so on. The last four scenarios depict those with interaction effects between mediators, with scenario 13 illustrating a setting with no residual correlation and scenario 16 illustrating a setting with a residual correlation level of 0.75}
\end{figure}

\begin{figure}[H]
\begin{center}
\includegraphics[width=0.7\textwidth]{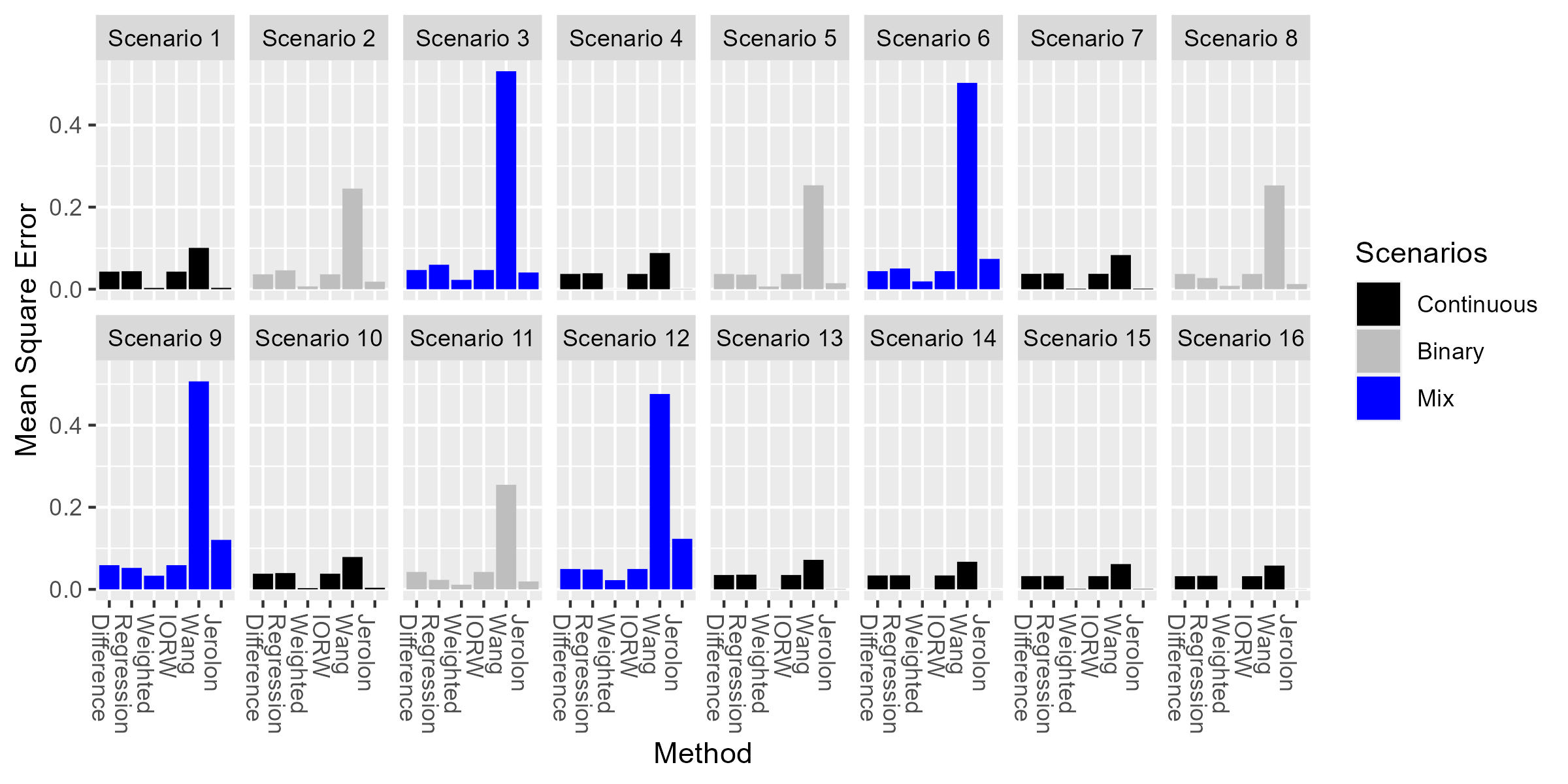}
\end{center}
\caption{Mean Square Error for TE across various scenarios as shown in \ref{scenario} The scenarios are arranged such that residual correlation increases from left to right: scenarios 1-3 have no residual correlation, scenarios 4-6 have a residual correlation of 0.25, and so on. The last four scenarios depict those with interaction effects between mediators, with scenario 13 illustrating a setting with no residual correlation and scenario 16 illustrating a setting with a residual correlation level of 0.75}
\end{figure}

\begin{figure}[H]
\begin{center}
\includegraphics[width=0.7\textwidth]{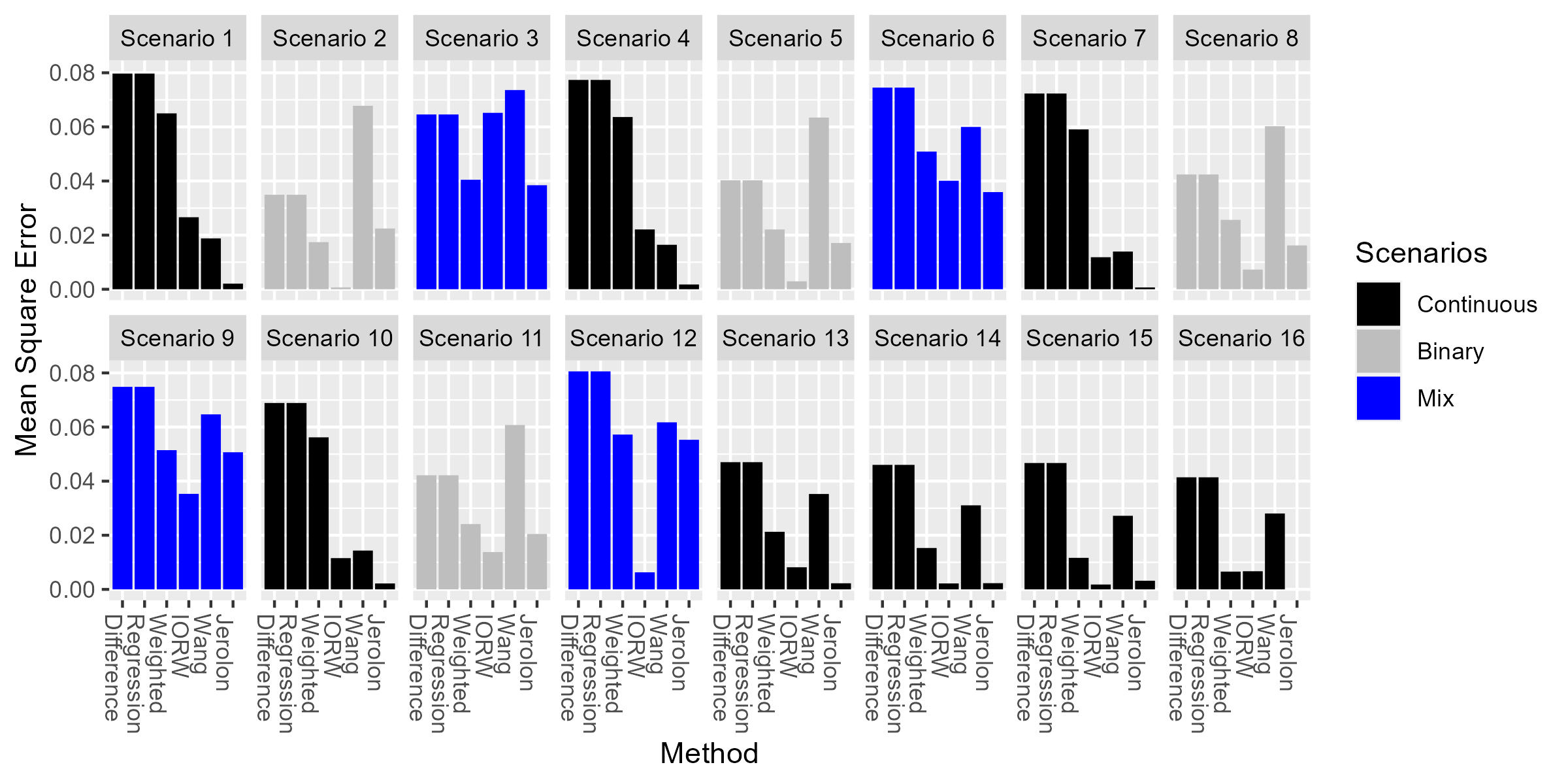}
\end{center}
\caption{Mean Square Error for DE across various scenarios as shown in \ref{scenario} The scenarios are arranged such that residual correlation increases from left to right: scenarios 1-3 have no residual correlation, scenarios 4-6 have a residual correlation of 0.25, and so on. The last four scenarios depict those with interaction effects between mediators, with scenario 13 illustrating a setting with no residual correlation and scenario 16 illustrating a setting with a residual correlation level of 0.75}
\end{figure}

\begin{figure}[H]
\begin{center}
\includegraphics[width=0.7\textwidth]{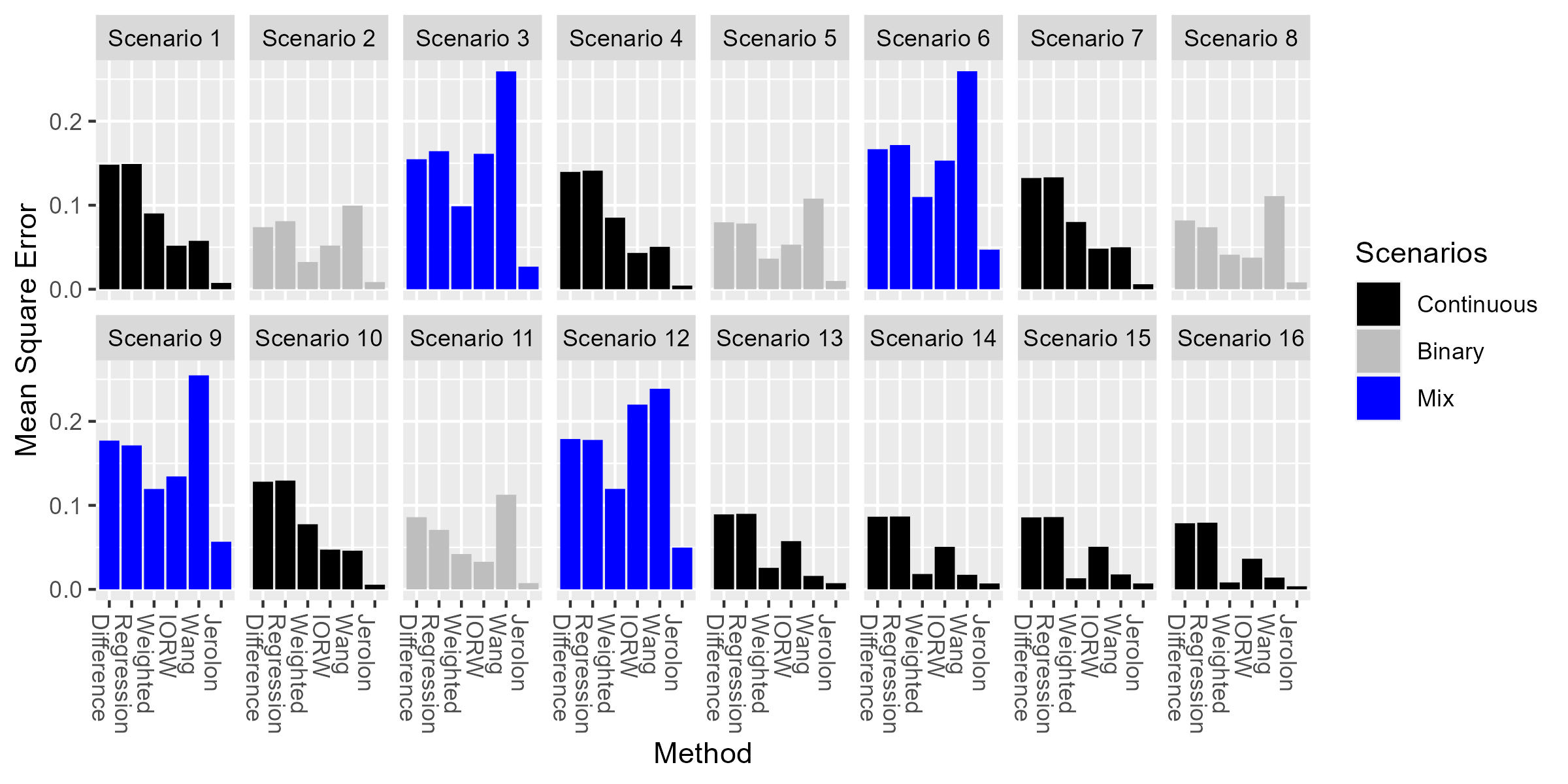}
\end{center}
\caption{Mean Square Error for IE across various scenarios as shown in \ref{scenario} The scenarios are arranged such that residual correlation increases from left to right: scenarios 1-3 have no residual correlation, scenarios 4-6 have a residual correlation of 0.25, and so on. The last four scenarios depict those with interaction effects between mediators, with scenario 13 illustrating a setting with no residual correlation and scenario 16 illustrating a setting with a residual correlation level of 0.75}
\end{figure}

\begin{figure}[H]
\begin{center}
\includegraphics[width=0.7\textwidth]{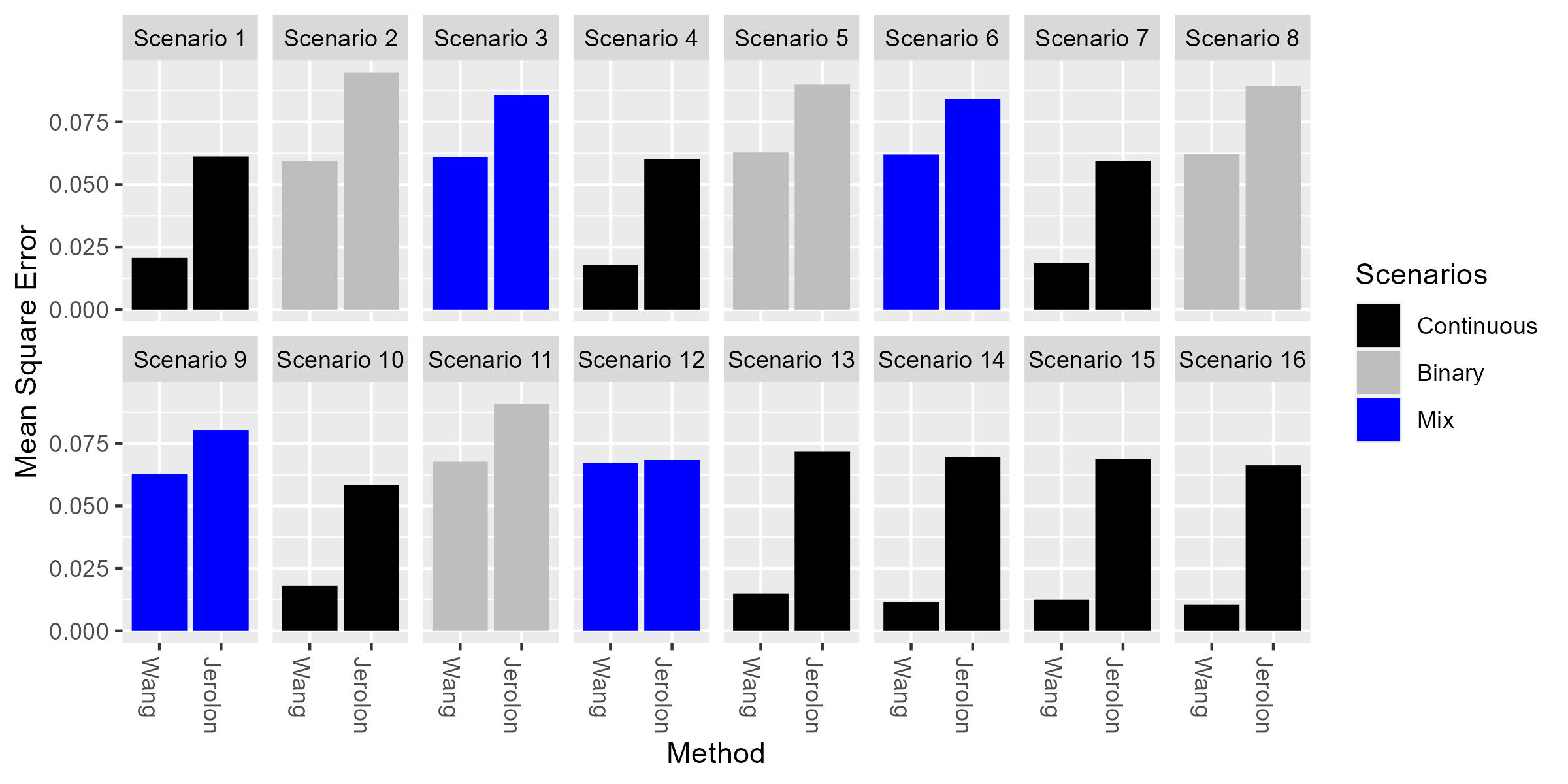}
\end{center}
\caption{Mean Square Error for IE for Mediator 1 across various scenarios as shown in \ref{scenario} The scenarios are arranged such that residual correlation increases from left to right: scenarios 1-3 have no residual correlation, scenarios 4-6 have a residual correlation of 0.25, and so on. The last four scenarios depict those with interaction effects between mediators, with scenario 13 illustrating a setting with no residual correlation and scenario 16 illustrating a setting with a residual correlation level of 0.75}
\end{figure}

\begin{figure}[H]
\begin{center}
\includegraphics[width=0.7\textwidth]{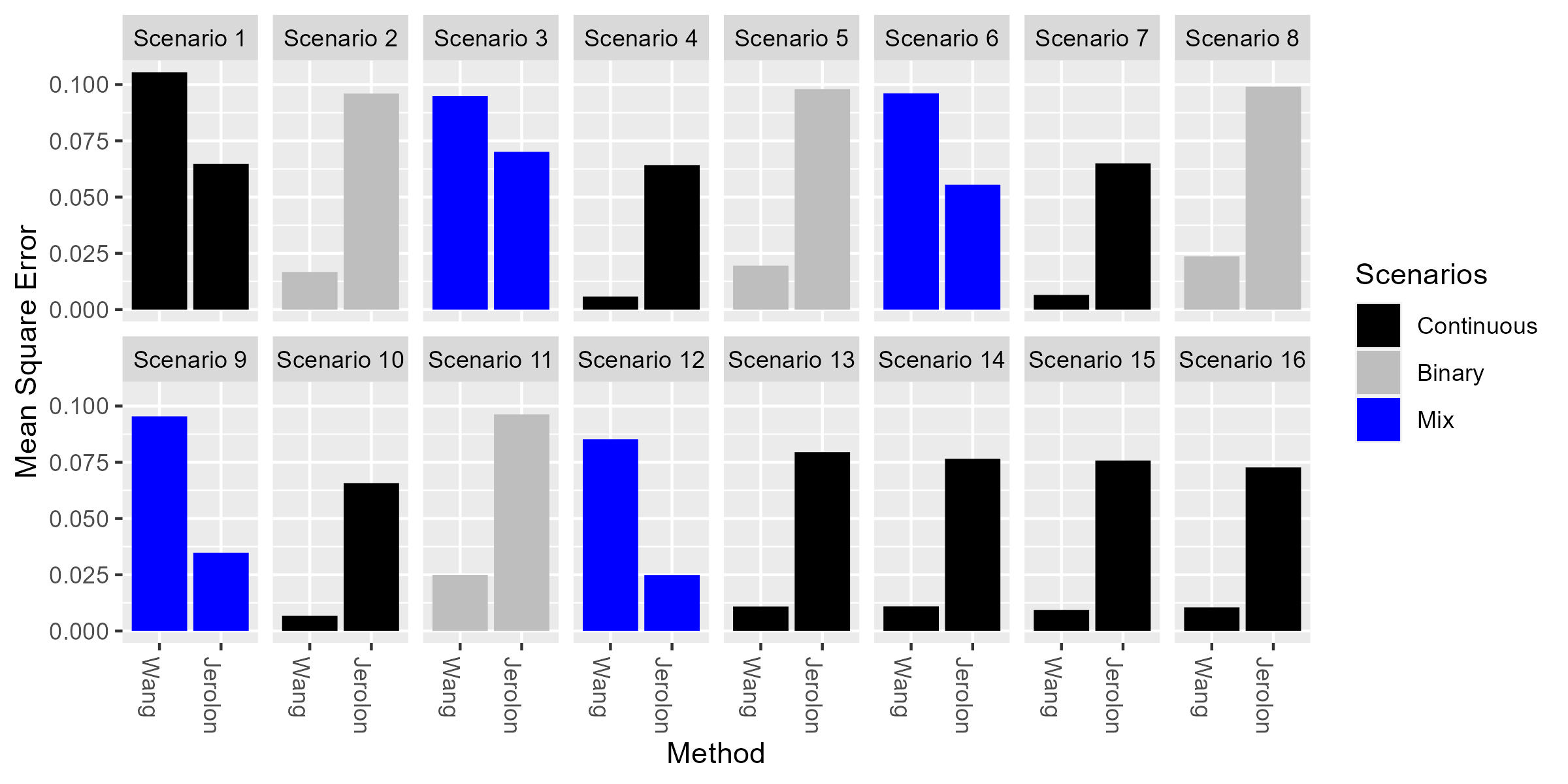}
\end{center}
\caption{Mean Square Error for IE for Mediator 2 across various scenarios as shown in \ref{scenario} The scenarios are arranged such that residual correlation increases from left to right: scenarios 1-3 have no residual correlation, scenarios 4-6 have a residual correlation of 0.25, and so on. The last four scenarios depict those with interaction effects between mediators, with scenario 13 illustrating a setting with no residual correlation and scenario 16 illustrating a setting with a residual correlation level of 0.75}
\end{figure}

\begin{figure}[H]
\begin{center}
\includegraphics[width=0.7\textwidth]{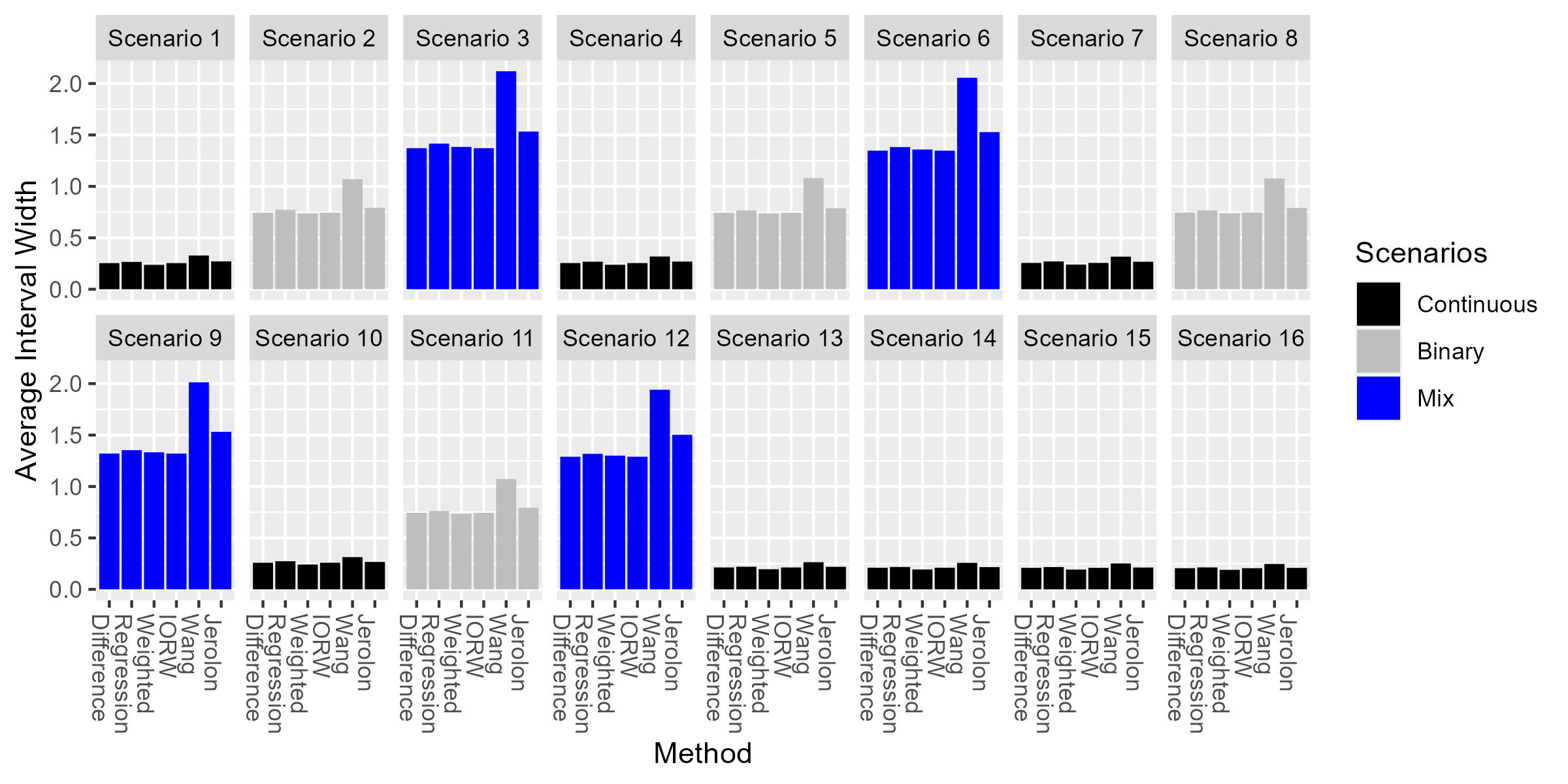}
\end{center}
\caption{Interval width of  TE across various scenarios as shown in \ref{scenario} The scenarios are arranged such that residual correlation increases from left to right: scenarios 1-3 have no residual correlation, scenarios 4-6 have a residual correlation of 0.25, and so on. The last four scenarios depict those with interaction effects between mediators, with scenario 13 illustrating a setting with no residual correlation and scenario 16 illustrating a setting with a residual correlation level of 0.75}
\end{figure}

\begin{figure}[H]
\begin{center}
\includegraphics[width=0.7\textwidth]{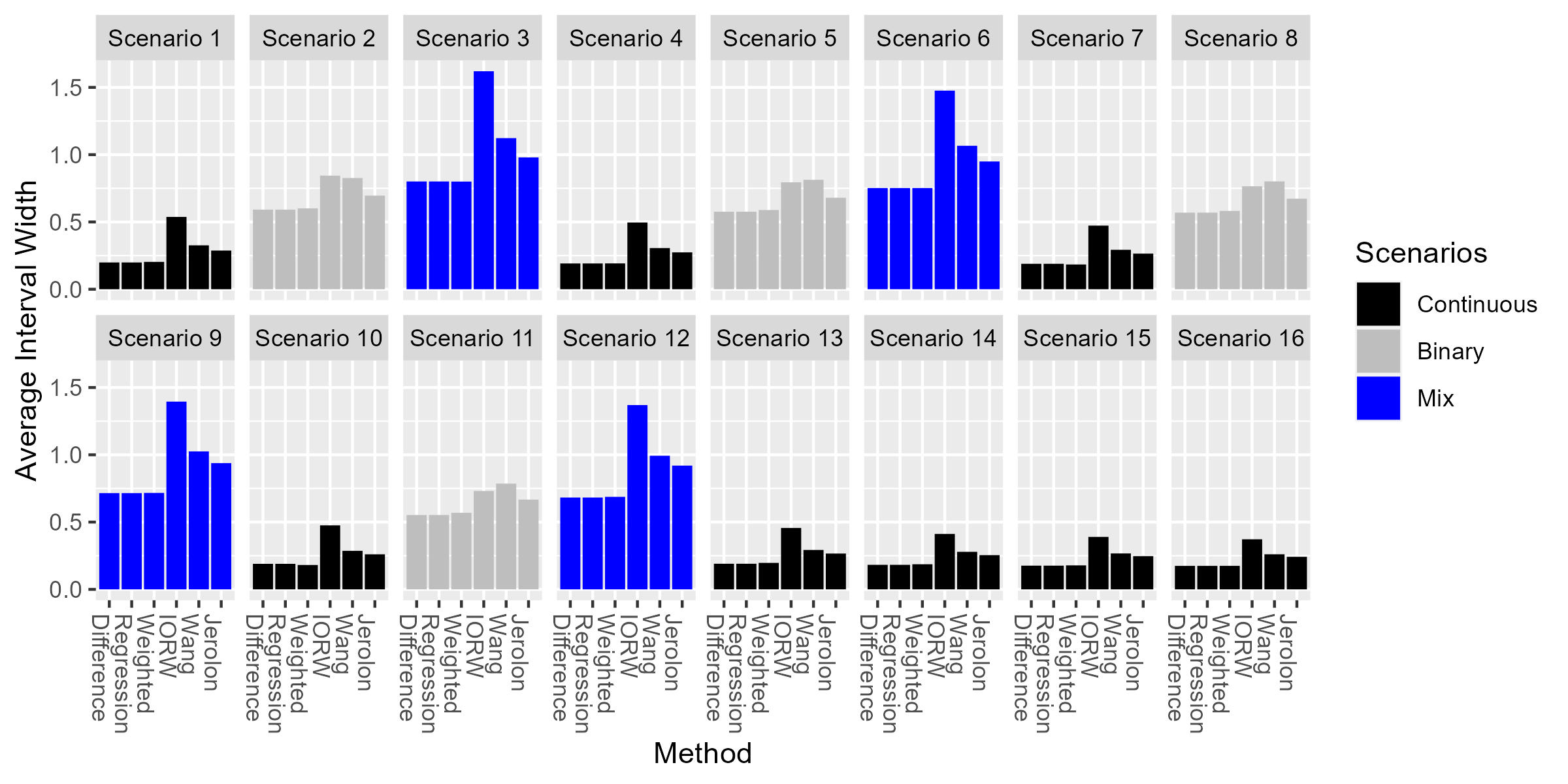}
\end{center}
\caption{Interval width of  DE across various scenarios as shown in \ref{scenario} The scenarios are arranged such that residual correlation increases from left to right: scenarios 1-3 have no residual correlation, scenarios 4-6 have a residual correlation of 0.25, and so on. The last four scenarios depict those with interaction effects between mediators, with scenario 13 illustrating a setting with no residual correlation and scenario 16 illustrating a setting with a residual correlation level of 0.75}
\end{figure}

\begin{figure}[H]
\begin{center}
\includegraphics[width=0.7\textwidth]{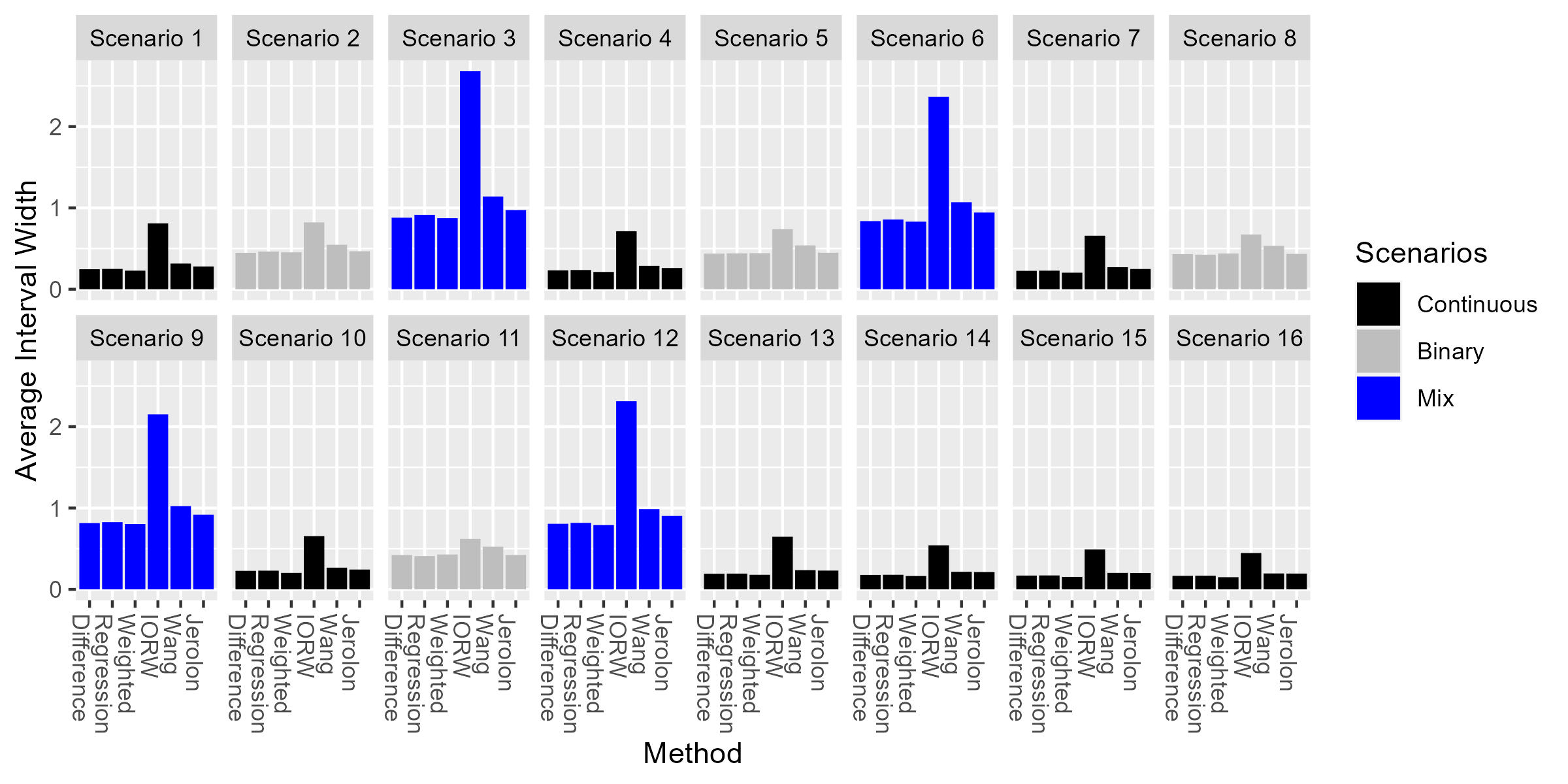}
\end{center}
\caption{Interval width of  IE across various scenarios as shown in \ref{scenario} The scenarios are arranged such that residual correlation increases from left to right: scenarios 1-3 have no residual correlation, scenarios 4-6 have a residual correlation of 0.25, and so on. The last four scenarios depict those with interaction effects between mediators, with scenario 13 illustrating a setting with no residual correlation and scenario 16 illustrating a setting with a residual correlation level of 0.75}
\end{figure}

\begin{figure}[H]
\begin{center}
\includegraphics[width=0.7\textwidth]{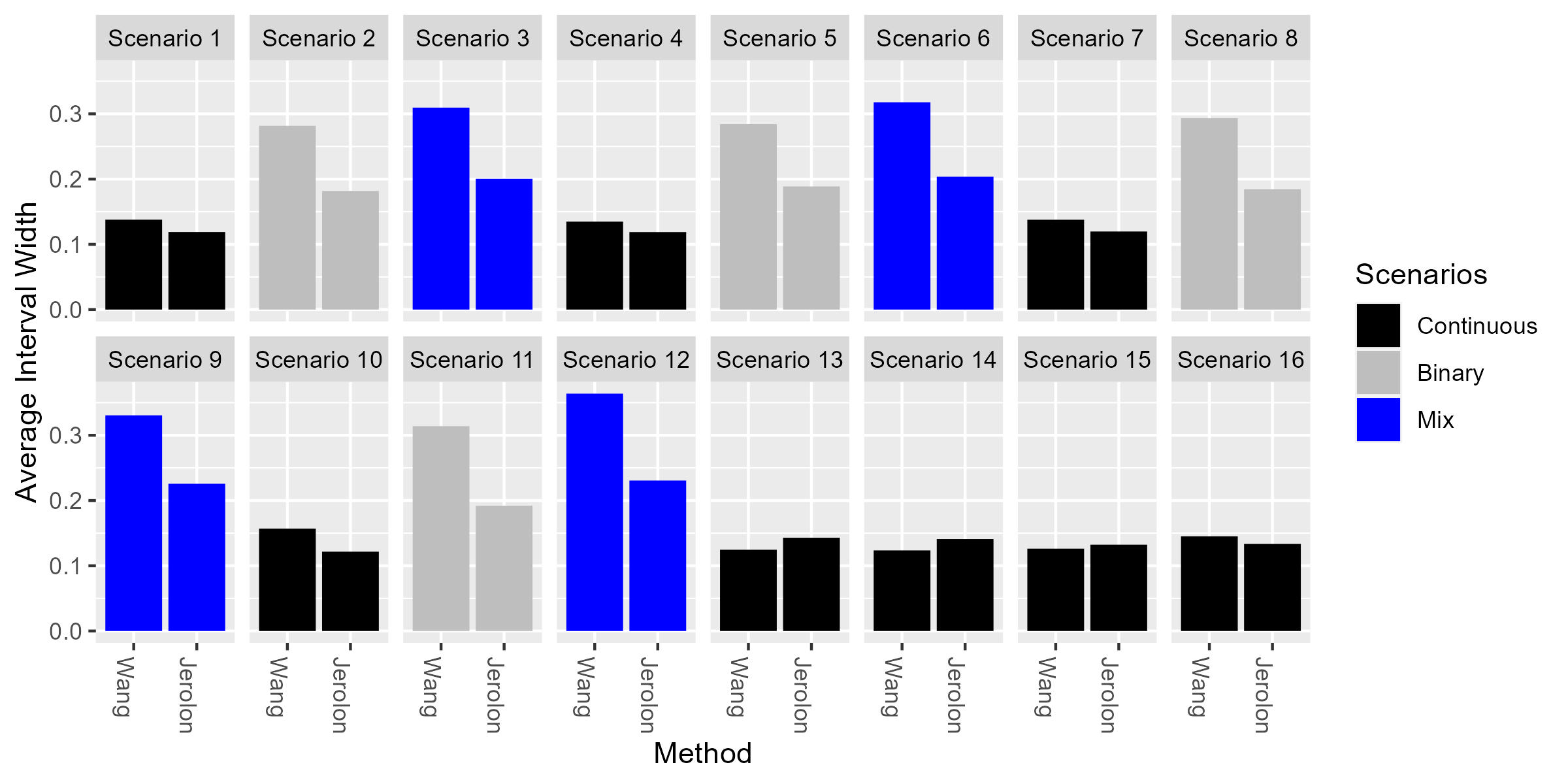}
\end{center}
\caption{Interval width of IE for mediator 1 across various scenarios as shown in \ref{scenario} The scenarios are arranged such that residual correlation increases from left to right: scenarios 1-3 have no residual correlation, scenarios 4-6 have a residual correlation of 0.25, and so on. The last four scenarios depict those with interaction effects between mediators, with scenario 13 illustrating a setting with no residual correlation and scenario 16 illustrating a setting with a residual correlation level of 0.75}
\end{figure}

\begin{figure}[H]
\begin{center}
\includegraphics[width=0.7\textwidth]{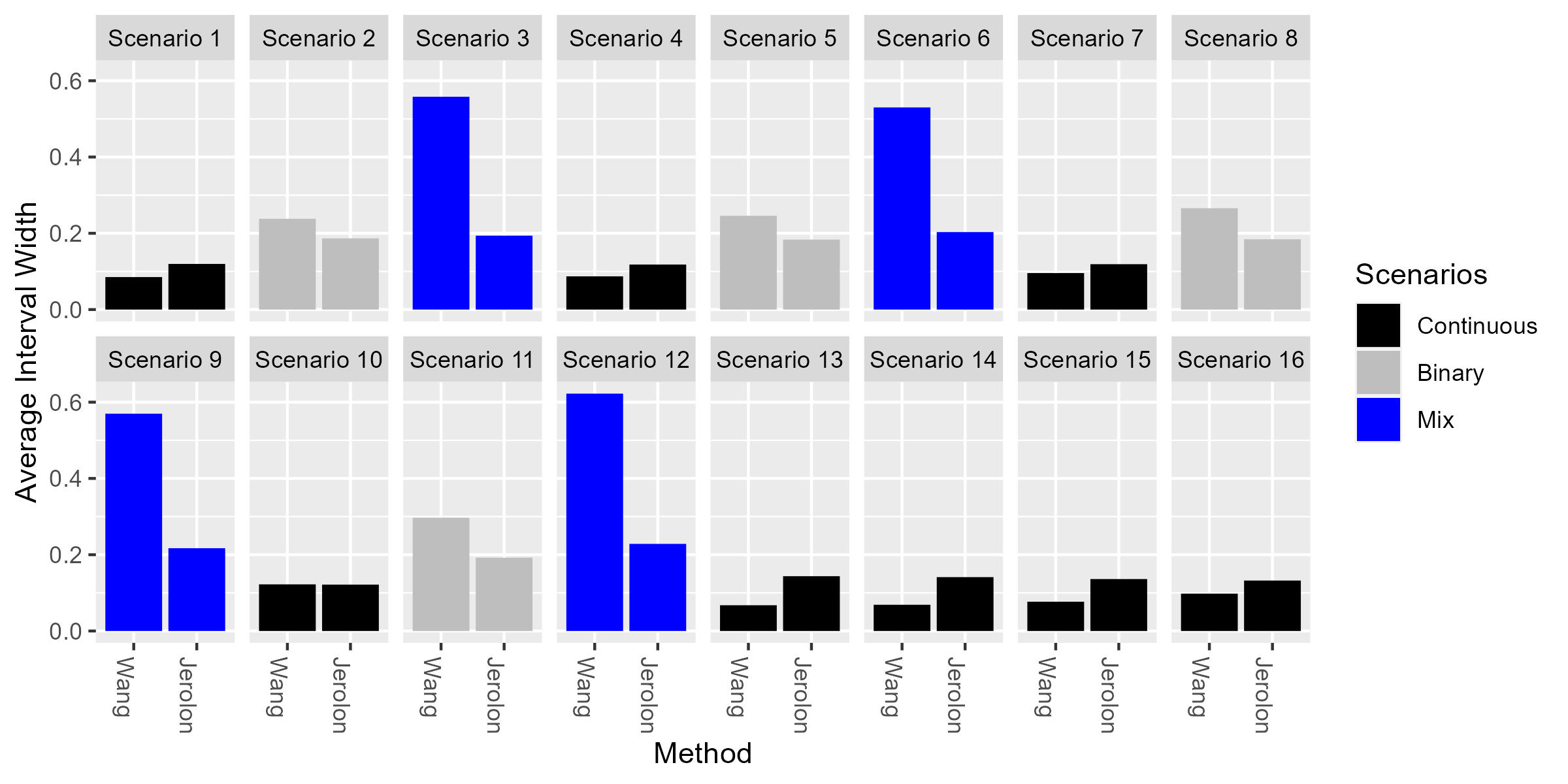}
\end{center}
\caption{Interval width of IE for mediator 2 across various scenarios as shown in \ref{scenario} The scenarios are arranged such that residual correlation increases from left to right: scenarios 1-3 have no residual correlation, scenarios 4-6 have a residual correlation of 0.25, and so on. The last four scenarios depict those with interaction effects between mediators, with scenario 13 illustrating a setting with no residual correlation and scenario 16 illustrating a setting with a residual correlation level of 0.75}
\end{figure}

\end{document}